\documentclass[%
reprint,https://www.overleaf.com/project/5d043a5f1813395b83bf03ba
,superscriptaddress,
 amsmath,amssymb,
 aps,
floatfix,
nofootinbib]{revtex4-1}

\usepackage[pdftex]{graphicx}
    \graphicspath{{figures/}}
    \DeclareGraphicsExtensions{.pdf,.png}
\usepackage{dcolumn}
\usepackage{bm}
\usepackage{color}
\usepackage{subcaption}
\usepackage{siunitx}

\usepackage[dvipsnames,svgnames,x11names]{xcolor}
\usepackage[markup=underlined]{changes}

\definechangesauthor[color=orange]{HP}
\definechangesauthor[color=blue]{GK}
\definechangesauthor[color=olive]{ZD}
\definechangesauthor[color=orange]{Peter}

\usepackage{todonotes}
\makeatletter
\setremarkmarkup{\todo[color=Changes@Color#1!20,size=\scriptsize]{#1: #2}}
\makeatother

\begin{document}

\title{ Strongly coupled Yukawa trilayer liquid: Structure and dynamics }

\author{Hong Pan}
\affiliation{Department of Physics, Boston College, Chestnut Hill, Massachusetts, 02467, USA}

\author{Gabor J. Kalman}
\affiliation{Department of Physics, Boston College, Chestnut Hill, Massachusetts, 02467, USA}

\author{Peter Hartmann}
\affiliation{Institute for Solid State Physics and Optics, Wigner Research Centre for Physics, P.O.Box. 49, H-1525 Budapest, Hungary}
\affiliation{Center for Astrophysics, Space Physics and Engineering Research (CASPER), Baylor University, 100 Research Pkwy, Waco, Texas 76706, USA}

\author{Zolt\'an Donk\'o}
\affiliation{Institute for Solid State Physics and Optics, Wigner Research Centre for Physics, P.O.Box. 49, H-1525 Budapest, Hungary}

\date{\today}

\begin{abstract}
The equilibrium structure and the dispersion relations of collective excitations in trilayer Yukawa systems in the strongly coupled liquid regime are examined. The equilibrium correlations reveal a  variety of structures in the liquid phase, reminiscent of the corresponding structures in the solid phase. At small layer separation substitutional disorder becomes the governing feature. Theoretical dispersion relations are obtained by applying the Quasi-Localised Charge Approximation (QLCA) formalism, while numerical data are  generated by micro-canonical molecular dynamics (MD) simulations. The dispersions and polarizations of the collective excitations obtained through both of these methods are compared and discussed in detail. We find that the QLCA method is, in general, very satisfactory, but that there are phenomena not covered by the QLCA. In particular, by analyzing the dynamical longitudinal and transverse current fluctuation spectra  we discover the existence of a novel structure, not related to the collective mode spectra. This also provides a new insight into the long-standing problem of the gap frequency discrepancy, observed in  strongly coupled layered systems in earlier studies.
\end{abstract}

\maketitle

\section{\label{sec:intro} Introduction}

Many-particle systems with classical pairwise inter-particle interactions are fundamental model systems that are successfully used for the investigation of a wide variety of complex phenomena. Depending on the choice of the interaction, these systems can be related to different atomic or molecular materials in all possible (classical) phases. In the limit of weak interaction, where the thermal motion of the particles dominates the ideal gas model becomes valid. At the other extreme of strong interactions, the thermal motion becomes irrelevant and the crystalline solid phase can be reached, where localised particle oscillations and lattice phonons govern the dynamics. At intermediate interaction strength, where the potential energy originating from the inter-particle interaction becomes comparable or even larger than the kinetic energy of the random thermal motion but is not strong enough to form crystalline structures, the so called ``strongly coupled liquid'' phase \cite{Ichimaru2004} is realized.

Relevant to our studies are the strongly coupled plasmas, where the inter-particle interaction acting between electrically charged particles can be well approximated with the electrostatic Coulomb interaction. In cases, where a polarizable background is present, the bare Coulomb force is screened. This screening can be approximated by an exponentially decaying factor, as it was introduced by Debye and H\"uckel for electrolytes \cite{Debye1923} and adopted to a number of physical systems, such as  dusty plasmas and charged colloidal suspensions under the name ``Yukawa potential'' \cite{Yukawa1935}. The advantage of such a  Yukawa One Component Plasma (YOCP) model system is that only one particle component has to be described explicitly, the contributions of all other constituents of a potentially complex system are subsumed by the modified interaction.

The popularity of the strongly coupled Yukawa model rapidly increased with the developments in the field of charged colloidal suspensions \cite{Winkle1988} and after the discovery of laboratory dusty plasmas and the realization of ``plasma crystals'' in 1994~\cite{Chu94,Thomas94,Melzer94}. Since that time a large variety of systems and phenomena have been investigated in great detail both experimentally and by means of numerical simulations. For a review of the early development in the field of dusty plasmas see e.g. Refs.~\cite{Fortov2004,Morfill2009,Bonitz2010}.

The interest in charged multi-layered systems started after pioneering works on Wigner crystals~\cite{Wigner1934} realized with electrons on the surface of superfluid He~\cite{Grimes1979}, semiconductor heterojunctions~\cite{Sarma1981}, and cold ions in traps~\cite{Mitchell1998,Bollinger2000}, which was followed by a series of theoretical studies on Coulomb systems~\cite{Vitk1984,Peeters1987,Falko1994,Bedanov1994,Esfarjani1995,Goldoni1995,Valtchinov1997,Kalman1999,kalman2000a,kalman2001,Ballester2003,Donko2003,Donko2003a,Ranganathan2008}. This line of research experienced a further boost after the discovery that macroscopic charged particle ensembles, such as colloids in a liquid suspension or dust particles in a gas discharge, when trapped in a narrow regions of space tend to self-organize into well distinguishable layers parallel to the boundaries instead of filling the volume homogeneously or in a close-packed configuration~\cite{Murray1990,Pieper1996,Neser1997,Zuzic2000,Teng2003,Hartmann2009,Reitmuller2013}. The more recent theoretical and numerical research triggered by these experiments has focused on the ground state structure~\cite{Totsuji1996,Totsuji1997,Messina2003,Mazars2008,Travenec2015,Donko2001,Golden2002} as well as on the collective excitations and instabilities~\cite{Golden2012,Ivlev17} in Yukawa bilayers. For multi-layered systems, the ground state structure~\cite{Bystrenko2003,Lowen2009,Contreras2010,Oguz2012} and its variation in the presence of shear and magnetic field~\cite{Lowen2005} have also been studied. Besides Yukawa systems, multi-layered configurations have been investigated in graphene~\cite{Yan2008,Cocemasov2013} and hard sphere systems~\cite{Sheu2008,Curk2012} as well.

The collective excitations of layered systems were first investigated by Das Sarma and Madhukar~\cite{Sarma1981}, who studied a Coulombic bilayer at weak coupling in the Random Phase Approximation (RPA) and predicted an out-of-phase acoustic excitation. Further studies revealed that at finite coupling the collective mode spectrum of layered systems is qualitatively different~\cite{Golden1993,Kalman1993,Ortner1999,Valtchinov1997,Kalman1999,kalman2000a}. Thus its analysis requires a theoretical approach incorporating strong coupling effects.

Two theoretical approaches have proven to be successful for the description and prediction of the collective mode structure and wave dispersion properties of classical charged particle systems in the strongly coupled liquid regime. Both, the Method of Moments~\cite{Krein1977,Igor1985,Igor2012, Arkhipov2017} and the Quasi-Localized Charge Approximation (QLCA)~\cite{Golden2000,Kalman2000,Kalman2005} connect the static structural properties of the particle ensembles to their dynamical, time dependent collective behavior. 

In this paper, we investigate the properties of a system where particles, interacting via the Yukawa potential, reside in three layers, constituting a ``Yukawa trilayer''. We focus on the strongly coupled liquid phase. In section~\ref{sec:modelsystem}  we describe the model system. The equilibrium structure of the system, based on the analysis of the pair distribution functions, is discussed  in section~\ref{sec:structure}. In section \ref{sec:QLCA} we invoke the QLCA theory and derive the wave dispersion relations. Amongst other things, we find a remarkable manifestation of the ``avoided crossing'' phenomenon, known to occur in a variety of other physical systems. Details of the Molecular Dynamics simulations are given in section \ref{sec:MD} and the results obtained through this method are presented and compared with the QLCA predictions in section \ref{sec:MDresults}. In section \ref{sec:envelope} we  address the issue of the unexpected anomalous behavior  of the system in the domain of small layer separations. Section \ref{sec:conclusion} gives a summary of our findings.

\section{\label{sec:modelsystem} Model system} 

We consider a system consisting of three identical 2D layers, infinite in the $x$ and $y$ directions, where the particles can move freely, with no displacement allowed in the $z$ direction. The layer in the center is labeled ``1'', while the layers at the top and bottom positions, respectively, are labeled  ``2'' and ``3''. The separation between the neighboring layers is $d$. The distance between any pair of layers is $s_{AB}$. Here, and in the sequel, $A,B$ indices are used to identify the layers. Each layer has an areal particle number density $n=n_{\rm s}/3$, where the $n_{\rm s}$ is the total projected surface particle density. We define the Wigner-Seitz (WS) radius $a$ based on the total density, i.e., $\pi a^2 n_{\rm s}=1$. In the sequel, distances, including the layer separation, are dimensionless, given in  units of $a$. Similarly, quantities of dimension of inverse distance, such as wave numbers, etc., are made dimensionless and are given in  units of $1/a$.

The particles interact through a pairwise Yukawa potential $\phi(r)=q\frac{{\rm e}^{-\kappa r}}{r}$, where $q$ is the particle charge (assumed to be the same for all particles), $\kappa$ is the screening parameter in units of $1/a$ and $r$ is the three-dimensional distance. Introducing the two-dimensional (projected onto the $x,y$ plane) distance $\rho$, the interaction potential energies between particles in layers $A$ and $B$, $\varphi^{AB}$, become: 

\begin{subequations}
\label{eq:yuka}
\begin{gather}
\varphi^{11}=\varphi^{22}=\varphi^{33} = q^2\frac{{\rm e}^{-\kappa \rho}}{\rho},  \\
\varphi^{12}=\varphi^{13} = q^2\frac{{\rm e}^{-\kappa \sqrt{\rho^2+d^2}}}{\sqrt{\rho^2+d^2}}, \\
\varphi^{23} = q^2\frac{{\rm e}^{-\kappa \sqrt{\rho^2+4d^2}}}{\sqrt{\rho^2+4d^2}}.
\end{gather}
\end{subequations}

The strength of Coulomb coupling is quantified with the coupling parameter: 
\begin{equation} \label{eq:Gamma}
    \Gamma=\beta q^2/a,
\end{equation}
where $\beta$ is $1/k_{\rm B}\,T$, and $k_{\rm B}$ is the Boltzmann constant. The characteristic frequency of the system is the 2D nominal plasma frequency $\omega_{\rm p}^2=2\pi q^2 n_{\rm s}/ma$, with $m$ denoting the uniform particle mass. In the following, frequencies are also made dimensionless and are given in units of $\omega_{\rm p}$.

The state and the behavior of a system defined above is completely determined by the three parameters: $\kappa$, $\Gamma$ and the inter-layer distance $d$.

The equilibrium  structure and the collective mode spectrum of the trilayer system is investigated here in the strongly coupled $\Gamma \gg 1$ liquid phase. The analysis is conducted for a Yukawa potential with $\kappa=0.4$. The system is described both via numerical simulations (based on the Molecular Dynamics (MD) approach) and theoretically by the Quasi-Localized Charge Approximation, which has been successfully used for a number of systems governed by Coulomb and Yukawa, as well as other types of  interactions~\cite{Rosenberg1997,Kalman2000,kalman2004,Hou2009,Khrapak2015}. The MD simulations yield both information about the structural properties and about the dynamical fluctuation spectra which also reveal the collective excitations, while the QLCA derives the mode structure and dispersion relations from the static data obtained in the simulations.


\section{\label{sec:structure} Structure}

\begin{figure}[htbp]
\begin{center}
   \includegraphics[width=0.7\columnwidth]{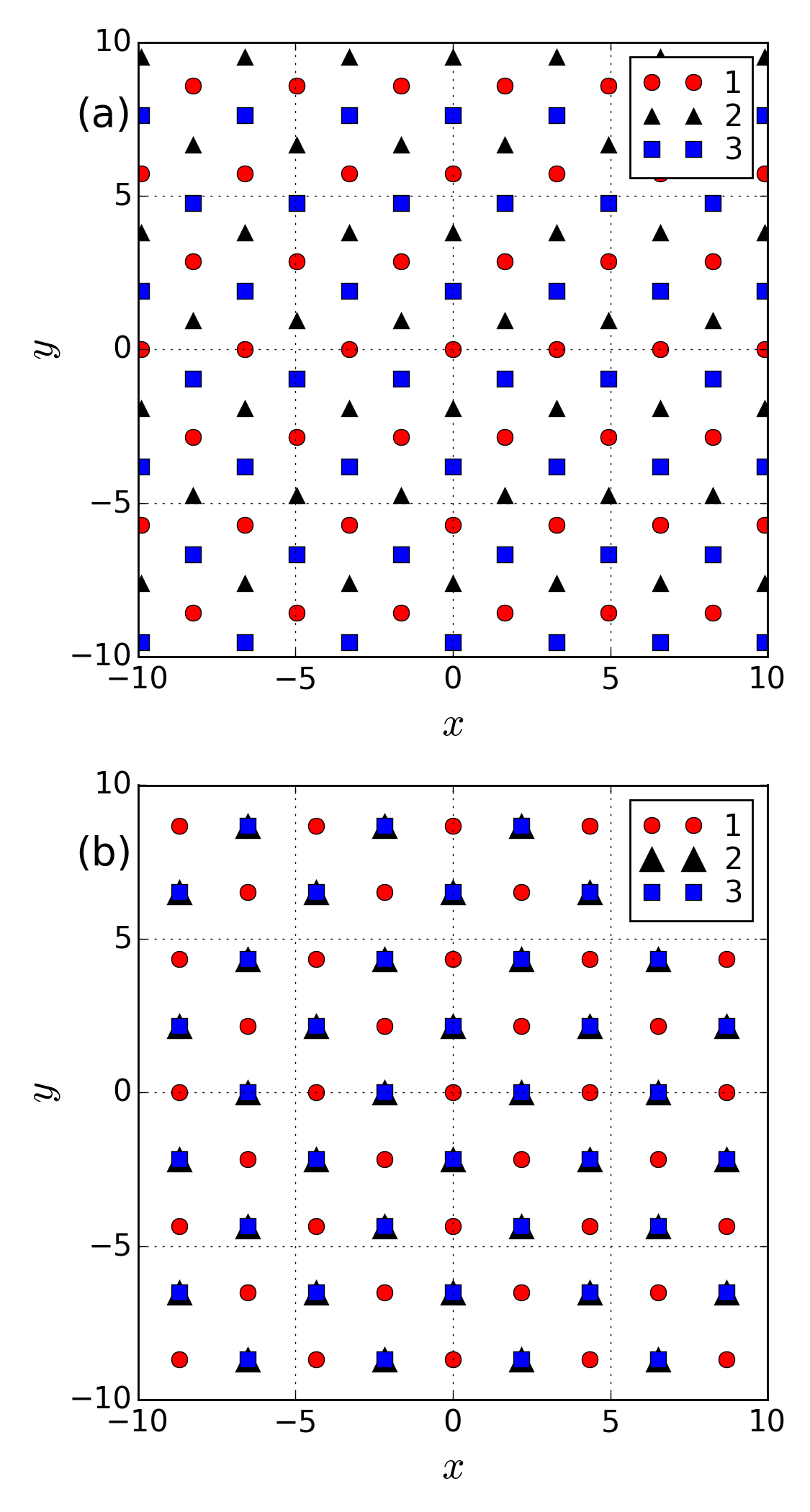}
   \caption{The two principal trilayer lattice structures (a) staggered hexagonal (SH) lattice, (b) overlapping square (OS) lattice. Note that here and in all figures in the sequel axis labels represent dimensionless quantities, as defined in the text.}
   \label{fig:osss}
\end{center}
\end{figure}

The equilibrium structure of a trilayer in the liquid state is well characterized by the set of pair correlation functions (PCF), $h_{AB}(\rho)$, or the equivalent pair distribution functions (PDF), $g_{AB}(\rho)=1+h_{AB}(\rho)$. (Note that the inter-layer correlation functions are given as functions of the projected distance $\rho$, rather than the actual distance $r$). At strong coupling the system shows a remarkable tendency to develop structures that vary dramatically with changing layer separation. The correlations in the liquid state are reminiscent of those characteristic for the underlying  structures that would form at the same $d$ value in the solid phase trilayer. In this respect, the trilayer is similar to the bilayer~\cite{Golden2002,Donko2001}, but for an additional degree of freedom that allows for two possible relative configurations (ABA and ABC stackings) of the top and bottom layers. The corresponding two of the principal crystal structures, as identified by~\cite{Lowen2009} are  the overlapping square (OS) and the staggered hexagonal (SH) lattices, illustrated in Fig.~\ref{fig:osss}. With varying layer separation other types of structures (not discussed here in details) are also realised.

\begin{figure}[htbp]
\begin{center}
    \includegraphics[width=0.9\columnwidth]{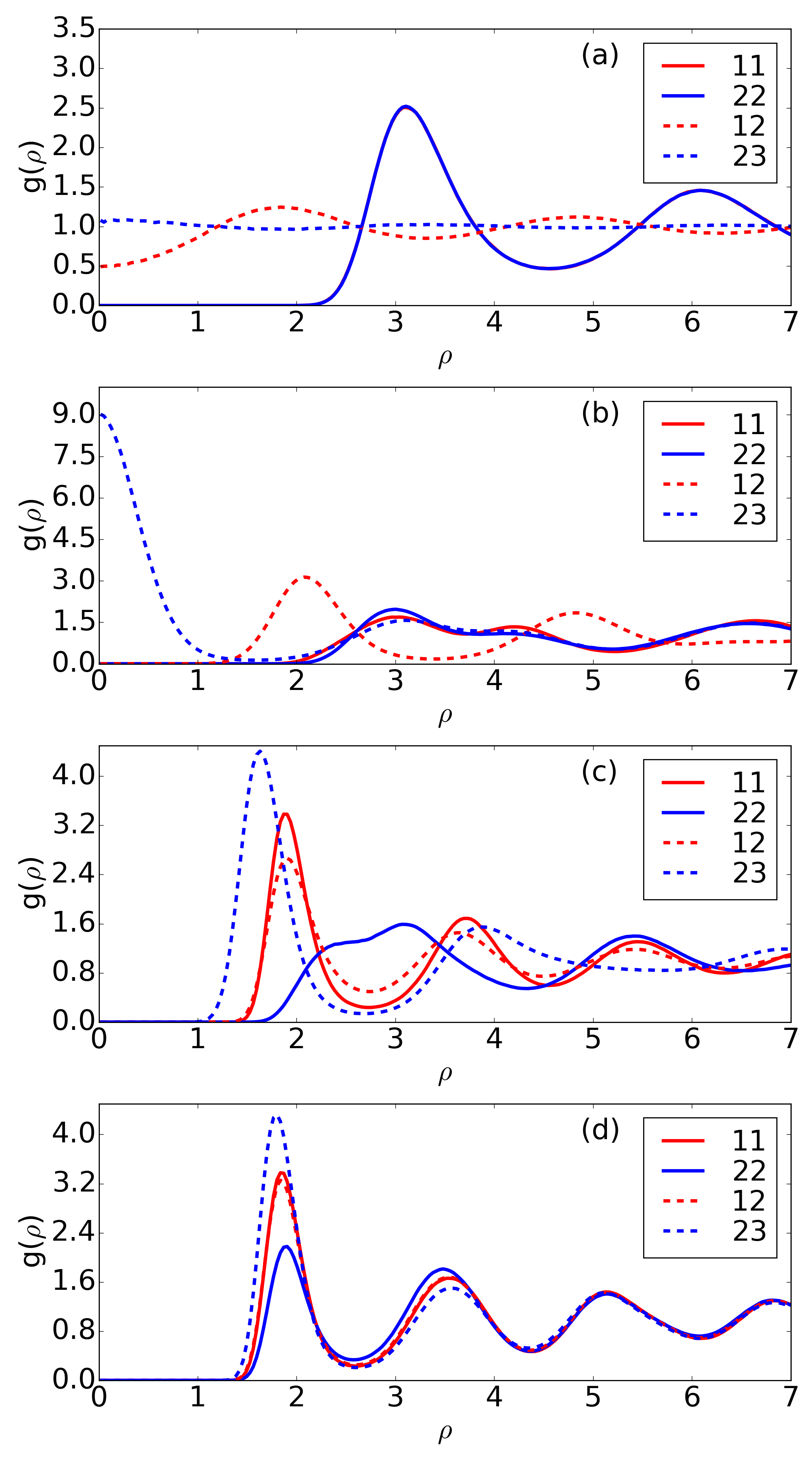}
    \caption{Intra-layer (11, 22) and inter-layer (12, 23) pair distribution functions for $\Gamma=160$, at layer separations (a) $d=3.0$, (b) $d=1.5$, (c) $d=0.5$, (d) $d=0.2$. Note that $g_{23}(\rho\to 0)>1$ at larger $d$ values and $g_{23}(\rho\to 0)<1$ at smaller $d$ values. In (a), $g_{11}(\rho)$ and $g_{22}(\rho)$ nearly overlap, while in (d) all $g(\rho)$-s nearly overlap apart from scale; see text for interpretation. Note that $g_{33}(\rho)$ is not shown as it is identical to $g_{22}(\rho)$ and $g_{13}(\rho)$ is not shown as it is identical to $g_{12}(\rho)$}
    \label{fig:gr1}
\end{center}
\end{figure}

\begin{figure}[htbp]
\begin{center}
    \includegraphics[width=0.7\columnwidth]{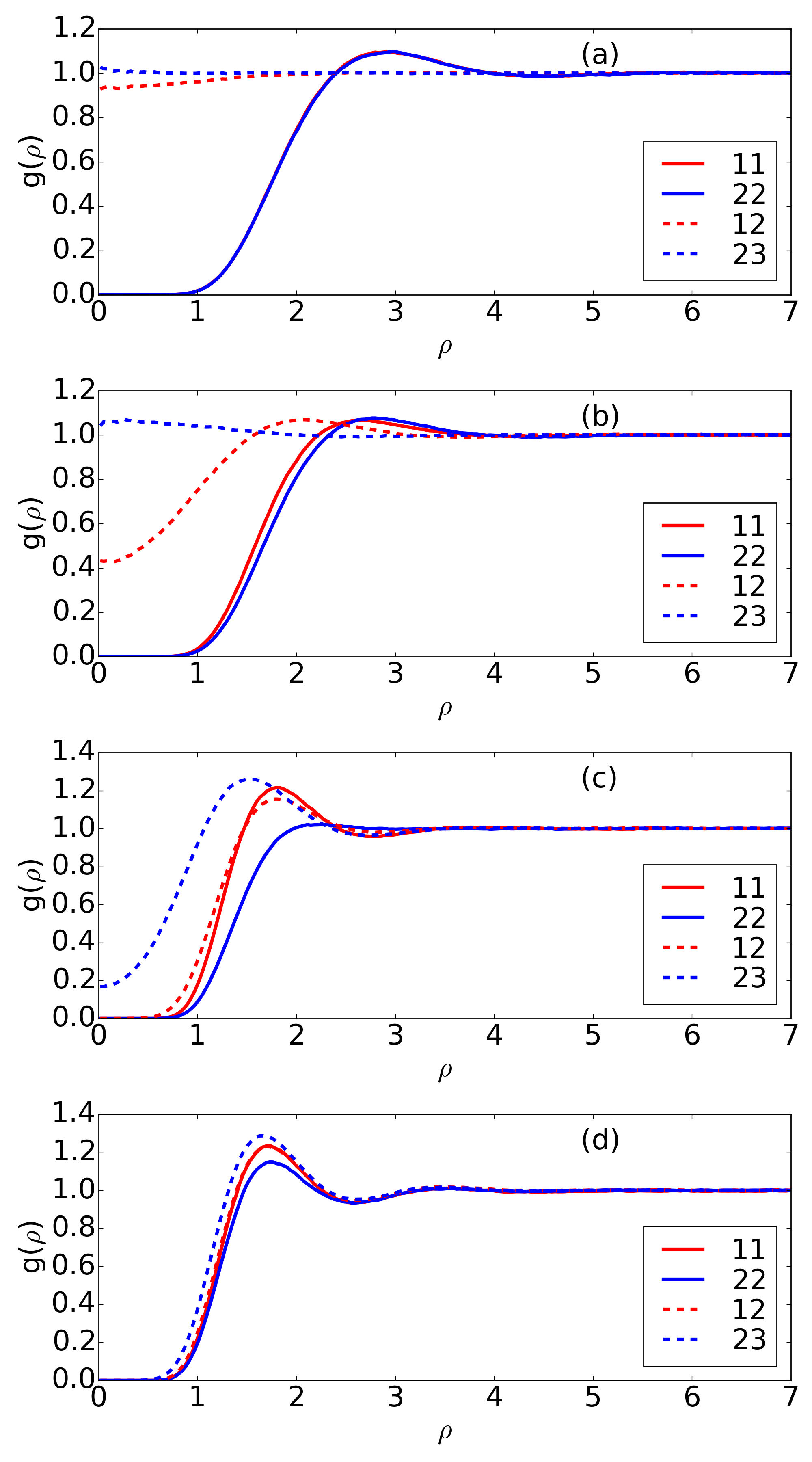}
    \caption{The same as Fig.~\ref{fig:gr1} for $\Gamma=10$. }
    \label{fig:gr2}
\end{center}
\end{figure}

The development of the correlations in the liquid phase is illustrated in Figs.~\ref{fig:gr1} and \ref{fig:gr2}, where a sequence of PDF-s at decreasing values of $d$ are displayed, for $\Gamma$ = 160 and 10, respectively. These figures present the intra-layer PDF-s $g_{11}(\rho)$ and $g_{22}(\rho)$, as well as the inter-layer PDF-s $g_{12}(\rho)$ and $g_{23}(\rho)$. The additional distribution functions $g_{33}(\rho)$ and $g_{13}(\rho)$ are not shown as these are identical to $g_{22}(\rho)$ and $g_{12}(\rho)$, respectively.

Focusing first on the high coupling, $\Gamma$ = 160 case, at the high $d=3$ value we observe a structure resembling the superposition of two bilayers~\cite{kalman2000a}, weakly correlated with each other. The lack of strong correlations is demonstrated by $g_{23}(\rho) \approx 1$. Also, even though layer 1, on one hand, and layers 2 and 3, on the other, are in different environments, this difference is not significant enough to induce a difference between $g_{11}(\rho)$ and $g_{22}(\rho)$: all the intra-layer PDF-s overlap and they exhibit correlations typical for an isolated 2D layer. The inter-layer PDF $g_{12}(\rho)$ shows that particle positions in layer 1 are staggered with respect to those in layers 2 and 3. These latter, in turn, trend to be positioned on top of each other in identical structures. These features are revealed by (i) $g_{12}(\rho=0)<1$, (ii) $g_{23}(\rho=0)>1$, and (iii) $g_{12}(\rho)$ exhibiting an out-of-phase and $g_{23}(\rho)$ exhibiting an in-phase behavior with respect to the extrema of $g_{11/22/33}(\rho)$. These features are indicative of an underlying OS crystal structure (see Figure~\ref{fig:osss}) of layers 2 and 3.

The correlations between the layers become much more emphasized as $d$ is decreased to $d=1.5$. Now, $g_{23}(\rho)$ assumes a very high peak at $\rho=0$, showing a strong overlap between the particle positions in layers 2 and 3. All this is consistent with a continued OS structure. The $g_{23}(\rho=0) \gg 1$ behavior may also be interpreted as an effective attraction between the top and bottom layers (mediated by the middle layer). 

Further decrease of $d$ to $d=0.5 $ changes the structure substantially. Now $g_{23}(\rho=0)=0$ and $g_{23}(\rho)$ is out-of-phase with respect to $g_{22}(\rho)$. These are signatures of the SH phase. However, the also visible strong separation of $g_{11}(\rho)$ and $g_{22}(\rho)$ and the development of an anomalous double-peak (i.e. two closely spaced peaks on a broad shoulder) in the latter are formations difficult to reconcile with either of the two simple lattice structures. These features seem to be related to the emergence of the so-called striped phase, a phase that has been encountered in many frustrated equilibria. Here the frustration is due to the competition between the direct interaction between layers 2 and 3, and their indirect interaction mediated through layer 1. The association of this domain with a striped phase is further suggested by the results of preliminary MD simulations on the solid phase of a trilayer, which indicate that such a striped phase is indeed  manifest there in the $1.0>d>0.5$ domain. The appearance striped phases have been reported during the melting of 2D crystals in \cite{Derzhko2009} and near the critical density in 2D and quasi-2D Coulomb systems in \cite{Reza2005}.

\begin{figure}[htbp]
\begin{center}
    \includegraphics[width=\columnwidth]{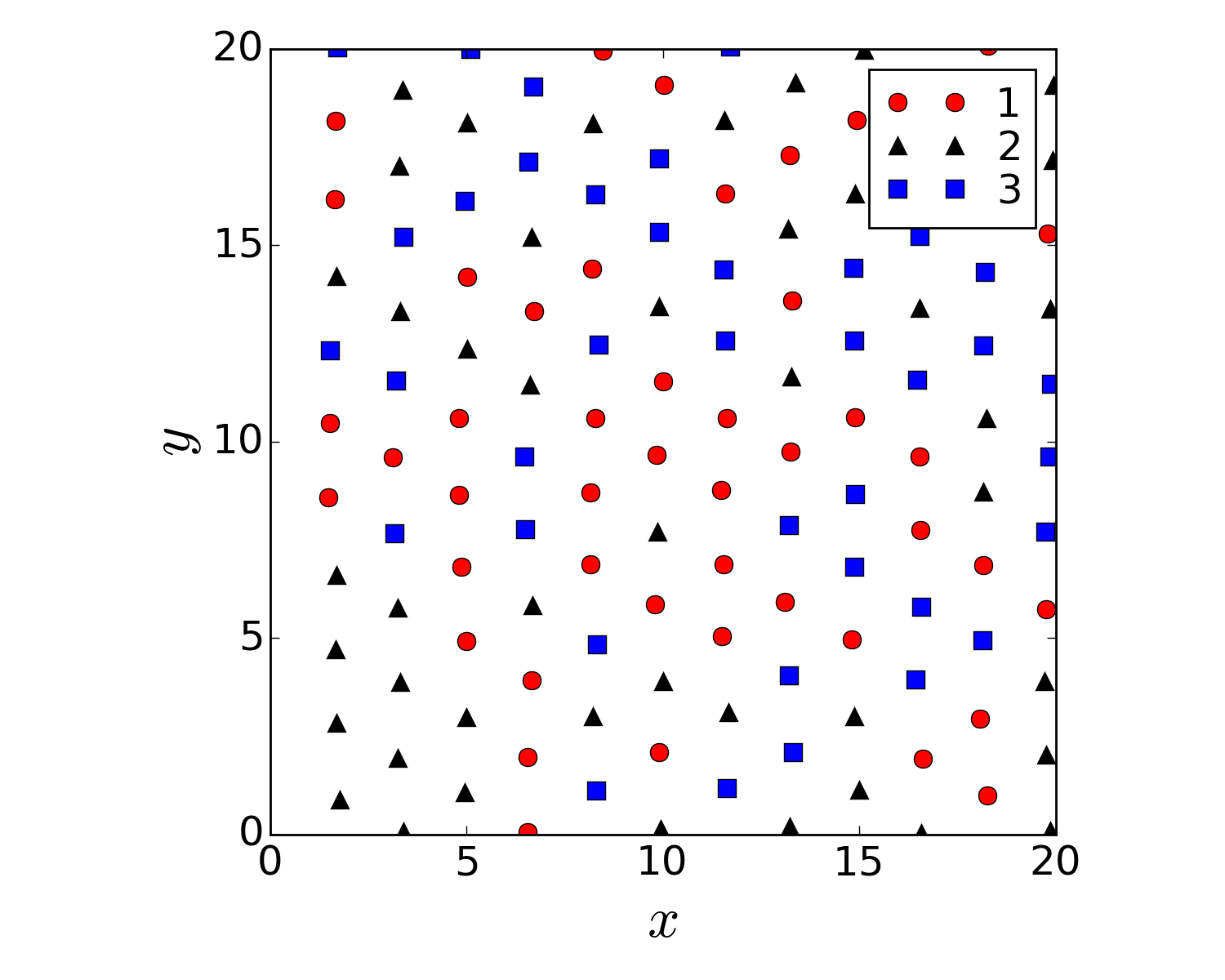}
    \caption{An overview  snapshot of the  trilayer system for  $\Gamma=1000$, at $d=0$ from MD simulation. The lattice structure is still nearly hexagonal, but the occupation of the lattice sites is random. Particle symbols indicate the actual layer, which is being occupied by the particle. This image is to be contrasted with panel (a) in Fig. \ref{fig:osss}, which depicts an ordered trilayer system. }
    \label{fig:disorder}
\end{center}
\end{figure}

\begin{figure}[htbp] 
\begin{center}
    \includegraphics[width=0.8\columnwidth]{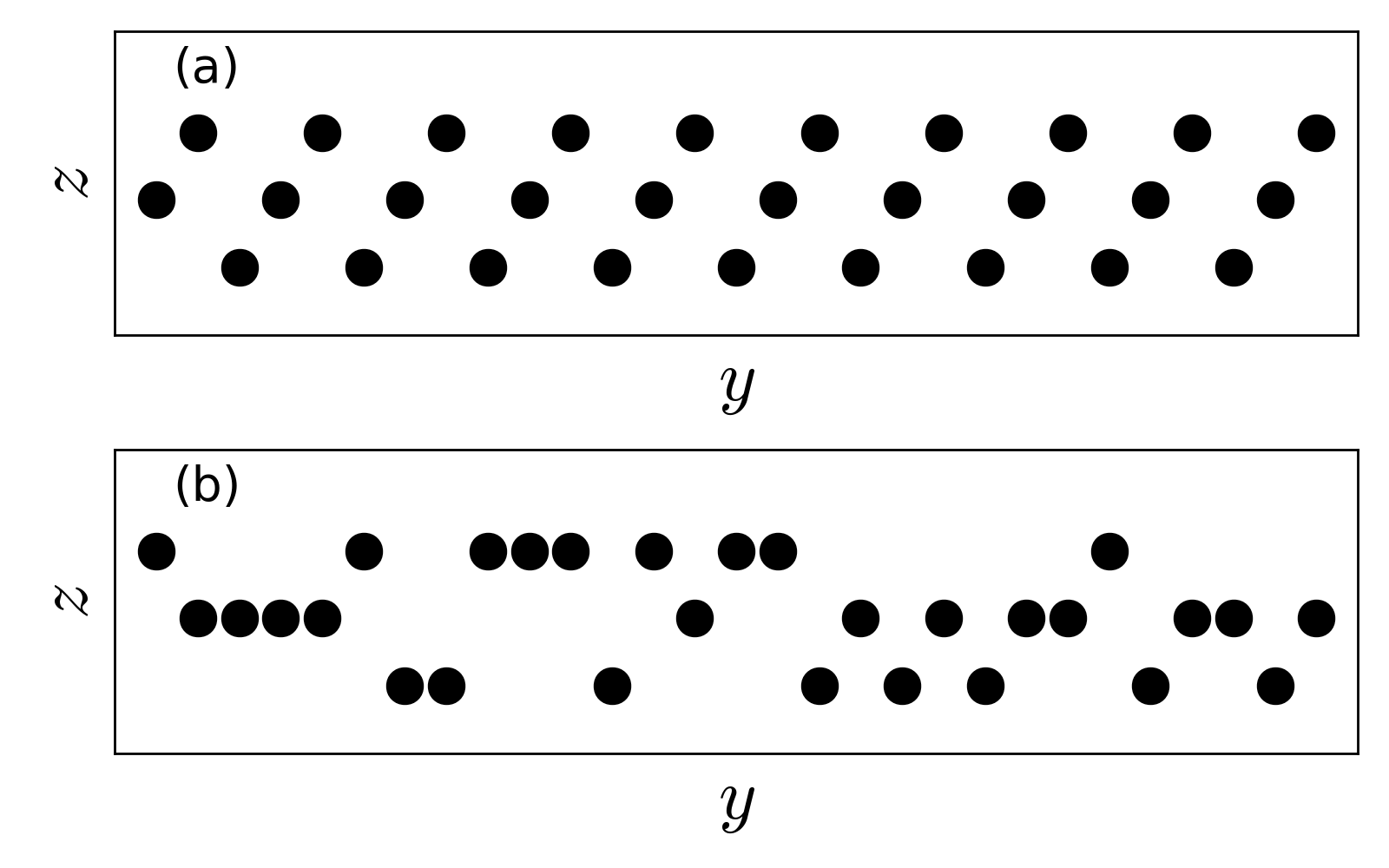}
    \caption{Side view of an arbitrarily chosen vertical row of particles in the SH crystal: (a) ordered, (b) substitutionally disordered. 
    }
    \label{fig:disorder1}
\end{center}
\end{figure}

Finally, when the layer distance is reduced to a very small value, $d=0.2$, an entirely new structure emerges: all the four PDF-s show the same behavior, only their peak amplitudes differ. This is the signature of the onset of a substitutional disorder. In this situation, the effect of the inter-layer separation on the inter-layer interaction becomes marginal, the particles tend to lose their layer identity and to occupy their sites in adjacent layers almost randomly. This structure is illustrated in Figs.~\ref{fig:disorder} and \ref{fig:disorder1}, where both the 2D projection from a high-$\Gamma$ simulation and a schematic side view of a disordered trilayer are portrayed. The substitutional disorder develops gradually with decreasing layer separation or with decreasing $\Gamma$ (cf. Figures~\ref{fig:disorder} and \ref{fig:disorder1}); when it is complete at $d \rightarrow 0$ the layers become statistically indistinguishable. The emergence of the substitutional disorder is the consequence of the requirement that at finite temperature (finite {$\Gamma$}) the Gibbs free energy of the system be minimized. A disordered state is obviously advantageous for maximizing the entropy, while the disadvantage it presents from the point of view of minimizing the interaction energy is small at low inter-layer separations~\cite{kalman2001}.

The PDF-s at the lower coupling value, $\Gamma$ = 10, displayed in Figure~\ref{fig:gr2}, reveal significantly weaker correlations, as expected. The trends, as functions of the layer separation, however, are similar to the high $\Gamma$ case, at $d$ = 3.0 (see Figure~\ref{fig:gr2}(a)) the particle positions in separate layers are almost uncorrelated: $g_{12}(\rho) \cong 1$ and $g_{23} (\rho)\cong 1$ for most $\rho$ values, except at small $\rho$. With decreasing $d$ there is a pronounced decay of $g_{12}(\rho)$ at small $\rho$ and in Figure~\ref{fig:gr2}(d), by reducing $d$ to smaller values we observe the onset of the substitutional disorder earlier.


\section{Collective mode spectrum: QLCA calculation} \label{sec:QLCA} 

We now proceed to the identification of the collective modes and to the derivation of their dispersion relations. As we have pointed out above, this is to be done within the framework of the QLCA. Following this approach, one has to calculate the dynamical matrix $C_{\mu\nu}^{AB}$. Here, $A$, $B$ are the layer indices and $\mu$, $\nu$ are the 2D Cartesian indices. For the system considered here $C_{\mu\nu}^{AB}$ is a $6\times6$ matrix,  the elements of which are expressed in terms of the $g_{AB}(\rho)$ PDF-s:
\begin{equation} \label{eq:mat1}
C_{\mu\nu}^{AB}(\vec{k})=\delta_{AB}\sum\limits_{D=1,2,3} M_{\mu\nu}^{AD}(0)-M_{\mu\nu}^{AB}(\vec{k}),
\end{equation}
where $\delta_{AB}$ is the Kronecker delta and 
\begin{equation} \label{eq:mat2a}
M_{\mu\nu}^{AB}(\vec{k})=\frac{q^2 n_s}{3ma}\int \psi_{\mu\nu}^{AB}(\vec{r}){\rm e}^{i\vec{k}\cdot\vec{r}} g_{AB}(\rho){\rm d}\vec{r}.
\end{equation}
Here, $q^2\psi_{\mu\nu}^{AB}(\vec{r})=\frac{\partial^2\varphi^{AB}(\vec{r})}{\partial r_{\mu}\partial r_{\nu}}$, thus
\begin{subequations}
\label{eq:cur5}
\begin{align}
\psi_{xx}^{AB} &=\frac{{\rm e}^{-\kappa r}}{r^5}[x^2(\kappa^2 r^2 +3\kappa r+3)-(1+\kappa r)r^2],  \\
\psi_{yy}^{AB} &=\frac{{\rm e}^{-\kappa r}}{r^5}[y^2(\kappa^2 r^2 +3\kappa r+3)-(1+\kappa r)r^2],  \\
\psi_{xy}^{AB} &=\frac{{\rm e}^{-\kappa r}}{r^5}[xy(\kappa^2 r^2 +3\kappa r+3)].
\end{align}
\end{subequations}

Due to the isotropy of the liquid phase the wave vector $\vec{k}$ can be fixed to $\vec{k} = (k,0)$, parallel to the $x$-axis, without the loss of generality and will be treated as a scalar variable in the following. Setting $r^2=\rho^2+s^2 $, where $\rho^2=x^2+y^2$, we have
\begin{equation} \label{eq:mat3}
M_{\mu\nu}^{AB}(k)=\frac{\omega_{\rm p}^2}{6\pi} \int_{0}^{\infty} \int_{0}^{2\pi} \psi_{\mu\nu}^{AB}(\rho,s,\theta){\rm e}^{ik\rho \cos\theta}
g_{AB}(\rho)\rho\, {\rm d}\rho {\rm d} \theta
\end{equation}

Performing first the angular integration, we define the kernel functions
\begin{equation} \label{eq:cur5a}
K_{\mu\nu}^{AB}(k\rho,r)=\frac{1}{6\pi}\int_{0}^{2\pi}\psi_{\mu\nu}^{AB}(\rho,s,\theta){\rm e}^{ik\rho \cos\theta} {\rm d} \theta. 
\end{equation}
Here, $x=\rho\cos\theta$, $y=\rho\sin\theta$. Note that in Eq.~\ref{eq:cur5} and in the sequel $s$ and $r$ are understood to depend on the indices $AB$. With the chosen direction of $\vec{k}$, reflection symmetry and the explicit $xy$ factor in Eq.~\ref{eq:cur5}c makes the integrals of the matrix element with $\psi_{xy} $ vanish. Defining also $P(\kappa r)=\kappa^2 r^2 +3\kappa r+3 $ and $Q(\kappa r)=1+\kappa r$ we arrive at 
\begin{eqnarray} \label{eq:mat4}
         K_{xx}^{AB}(k\rho,r) &=\frac{1}{6\pi}\frac{{\rm e}^{-\kappa r}}{r^5} \int_{0}^{2\pi}(P\rho^2 \cos^2 \theta-Qr^2) {\rm e}^{ik\rho \cos\theta} {\rm d} \theta  \nonumber \\
         &=\frac{{\rm e}^{-\kappa r}}{3r^5}\left[(\frac{J_1}{k\rho}-J_2)P\rho^2-Qr^2J_0\right]   \\
         K_{yy}^{AB}(k\rho,r) &=\frac{1}{6\pi}\frac{{\rm e}^{-\kappa r}}{r^5} \int_{0}^{2\pi}(P \rho^2 \sin^2 \theta-Qr^2) {\rm e}^{ik\rho \cos\theta} {\rm d} \theta  \nonumber \\
         &=\frac{{\rm e}^{-\kappa r}}{3r^5}\left[\frac{J_1}{k\rho}P\rho^2-Qr^2J_0\right].
\end{eqnarray}
The $J$-s are the respective  Bessel functions of $k\rho $. In the $k\to 0$ limit, these expressions reduce to 
\begin{equation} \label{eq:fxyf}
K_{xx}^{AB}(k\to 0)=K_{yy}^{AB}(k\to 0)=\frac{{\rm e}^{-\kappa r}}{3r^5} \left(\frac{1}{2}P\rho^2-Qr^2 \right).
\end{equation}
The elements of the $M$-matrix can now be expressed in terms of the integrals
\begin{equation}
    \label{eq:knint}
     M_{\mu\nu}^{AB}(k) =\omega^2_{\rm p}\int_{0}^{\infty} K_{\mu\nu}^{AB}(k\rho,r)g_{AB}(\rho)\rho {\rm d} \rho. 
\end{equation}
The dynamical matrix is a real symmetric matrix with real eigenvalues.  It is valued in the space which is the direct product of the 3D layer-space and the  2D Cartesian space. Due to the isotropy of the liquid state, when $\vec{k}$ points along the $x$-axis the matrix can be diagonalized into two diagonal, longitudinal (or $[xx]$) and transverse (or $[yy]$), $3 \times 3$ sub-matrices.

Both of the sub-matrices have the form 
\begin{equation}
\begin{pmatrix} 
E & Y & Y \\
Y & F & D \\
Y & D & F
\end{pmatrix}.
\end{equation}
The explicit expressions for the 8 distinct matrix elements are presented in the Appendix. 

A further unitary transformation, consisting of a $45^\circ$ rotation about the $1$-axis in layer space can cast the $C$-matrix into the semi-diagonal form
\begin{equation}\label{eq:MMM}
\begin{pmatrix} 
E & \sqrt{2}Y & 0 \\
\sqrt{2}Y & F+D & 0 \\
0 & 0 & F-D
\end{pmatrix}.
\end{equation}

\begin{figure}[htbp]
\begin{center}
\includegraphics[width=0.8\columnwidth]{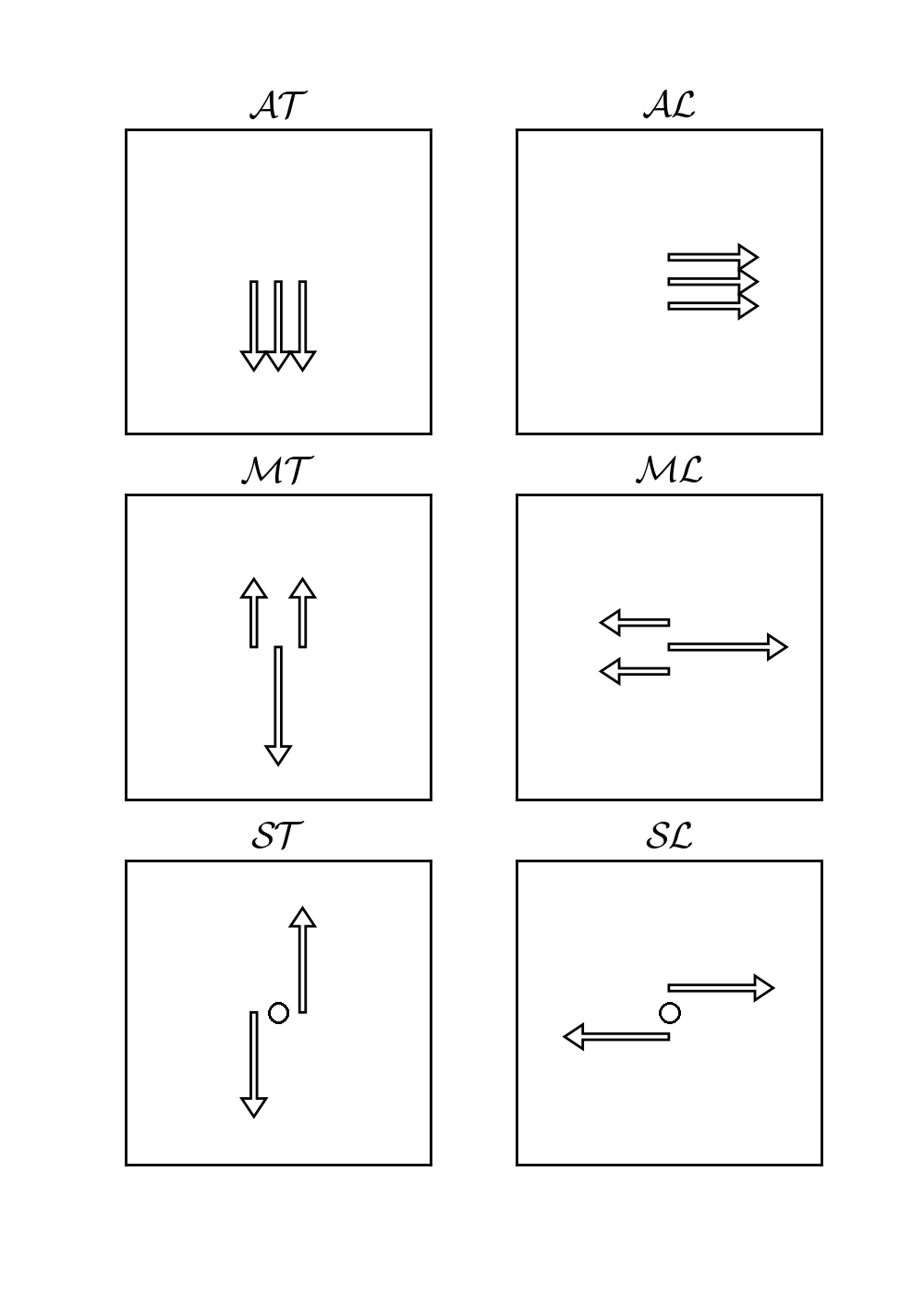}
\caption{Illustration of the 3 longitudinal and 3 transverse eigenvectors with $\vec{k}$ taken along the $x$ (horizontal) direction. Two sets of modes exist, each either with longitudinal ($\mathcal{L}$) or transverse ($\mathcal{T}$) Cartesian polarizations.  $\mathcal{A}$ mode: particles in all the three layers move in the same direction, $\mathcal{M}$ mode: particles in the middle layer (1) move in one direction, while those in layers 2 and 3 move together in the opposite direction, $\mathcal{S}$ mode: particles in the middle layer remain stationary, while those in layers 2 and 3 move in opposite directions with respect to each other. The arrows represent the displacements of the particles within the respective layers.}

\label{fig:po}
\end{center}
\end{figure}

Then the eigenvalues become
\begin{subequations} \label{eq:eigenvl}
\begin{align}
        \alpha_1=F-D,  \\
        \alpha_2=\frac{E+D+F-\sqrt{\Delta}}{2}, \\
        \alpha_3=\frac{E+D+F+\sqrt{\Delta}}{2},
        \end{align}
\end{subequations}
with their corresponding eigenvectors (in the original coordinate system)
\begin{subequations} \label{eq:eigenvc}
\begin{align}
        \vec{v}_1=(0,-1,1),  \\
        \vec{v}_2=\left(-\frac{-E+D+F+\sqrt{\Delta}}{2Y},1,1 \right),  \\
        \vec{v}_3=\left(-\frac{-E+D+F-\sqrt{\Delta}}{2Y},1,1 \right), 
        \end{align}
\end{subequations}
where $\Delta=(F+D-E)^2+8Y^2$. We also note that the  Einstein frequency $\Omega_A$ (the oscillation frequency of a particle in layer $A$ in the presence of the frozen environment of all the other particles) is given by $\Omega_A^2=\sum_{B=1,2,3}M_{xx}^{AB}(0)=\sum_{B=1,2,3}M_{yy}^{AB}(0)$. It should be kept in mind that the matrix elements in the above formulae are valued either in the longitudinal or in the transverse subspaces. Thus, in terms of their Cartesian polarizations there are three  longitudinal ($\mathcal{L}$) and three transverse ($\mathcal{T}$) eigenmodes. Out of each the three modes one is acoustic ($\mathcal{A}$), while the two remaining ones are gapped (optic) modes. Their polarizations in layer-space are distinguished by the relative displacements of the layers. In the $\mathcal{A}$ mode the particles in all the three layers move in the same direction; in one of the gapped modes, labelled $\mathcal{S}$, particles in the middle layer 1 remain stationary, while particles in layers 2 and 3 move in opposite directions with respect to each other; in the other gapped mode, labelled $\mathcal{M}$, particles in the middle layer 1 move in one direction, while those in layers 2 and 3 move together in the opposite direction. These polarizations are schematically illustrated in Figure~\ref{fig:po}. 

\begin{figure}[htbp]
\begin{center}
\includegraphics[width=\columnwidth]{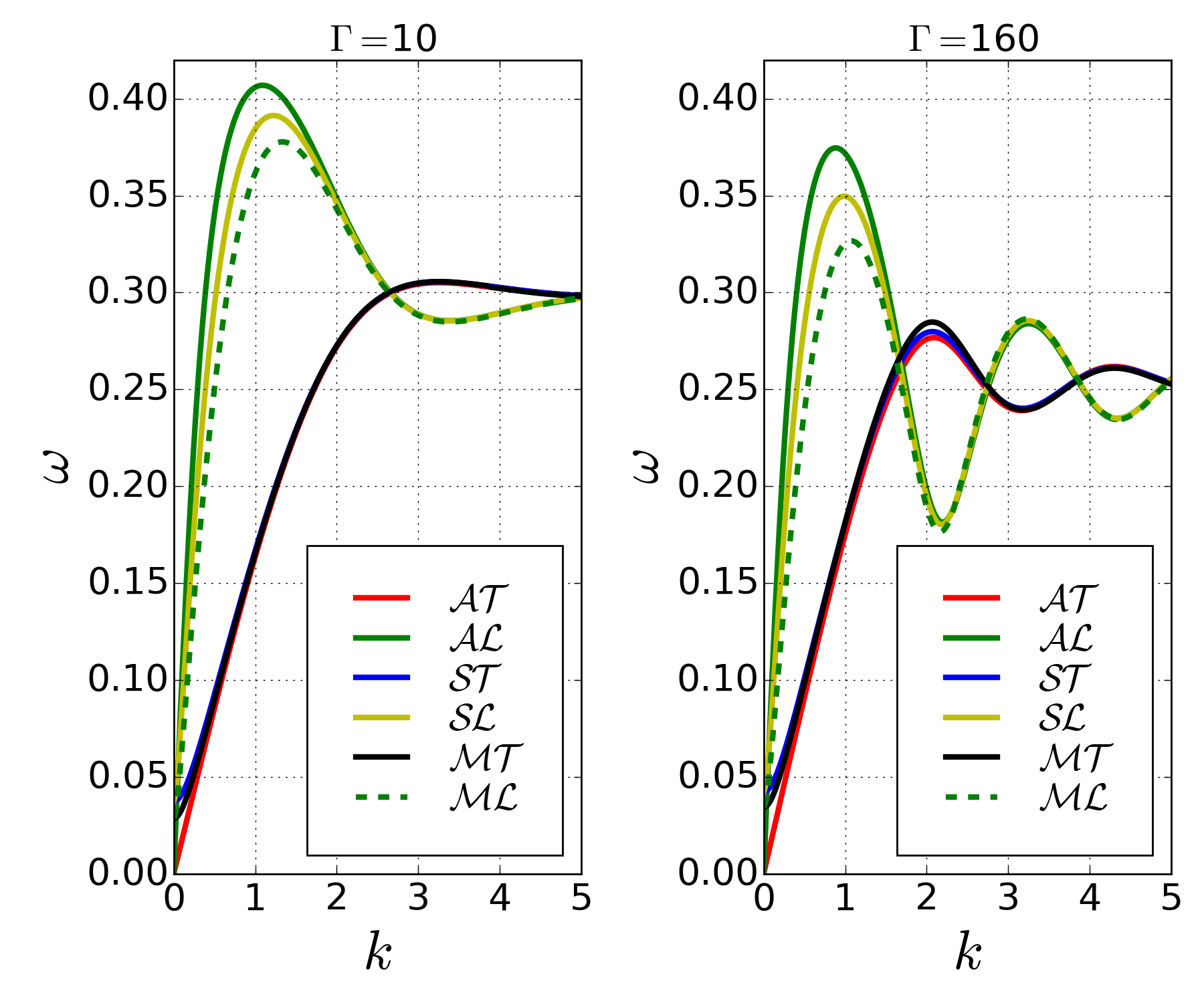}
\caption{QLCA dispersion curves of the six collective modes of the trilayer system, for moderate coupling $\Gamma=10$, and high coupling $\Gamma=160$ values, at $d=3.0$. }
\label{fig:qc1}
\end{center}
\end{figure}

\begin{figure}[htbp]
\begin{center}
\includegraphics[width=\columnwidth]{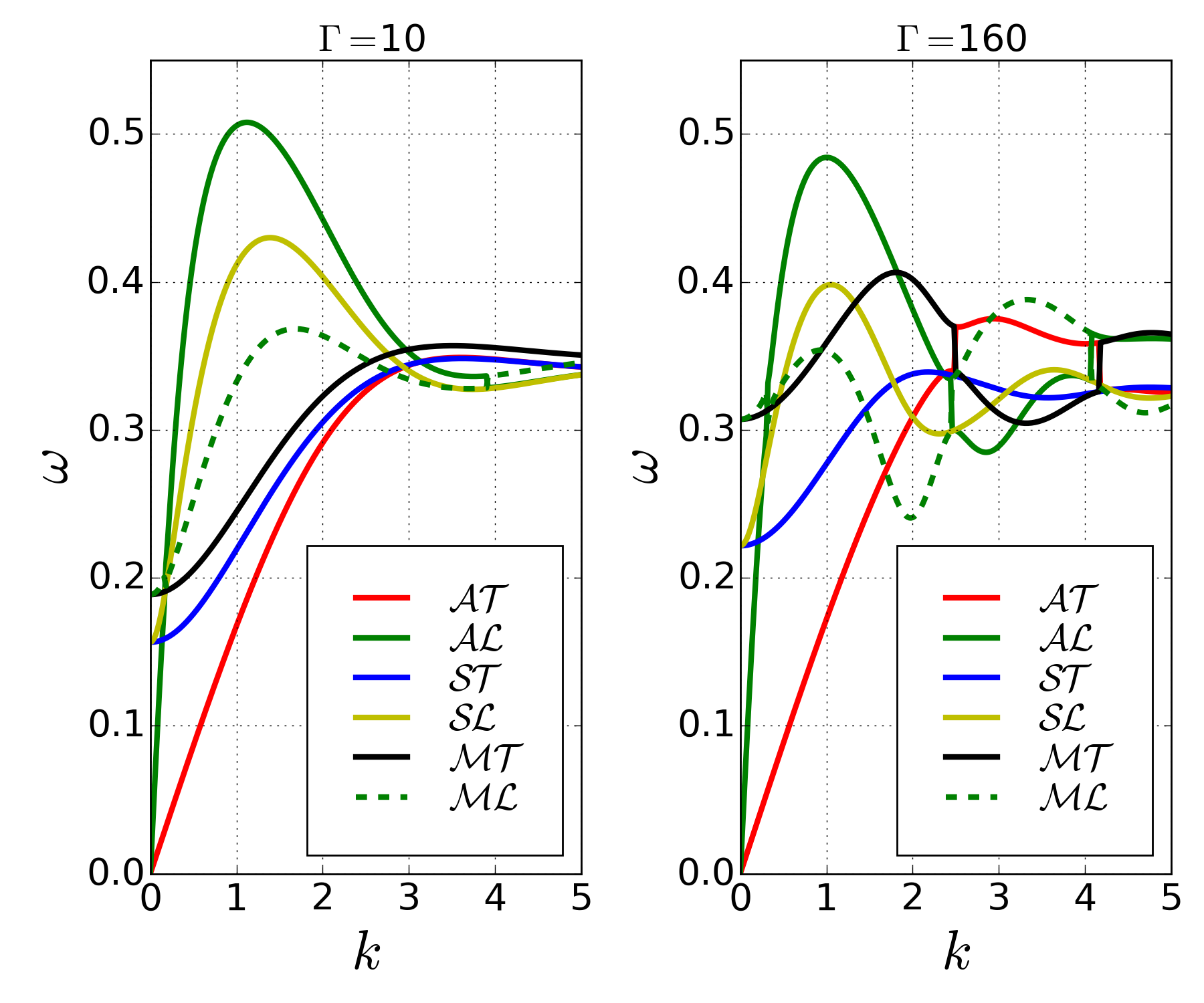}
\caption{QLCA dispersion curves of the six collective modes of the trilayer system, for moderate coupling $\Gamma=10$, and high coupling $\Gamma=160$ values, at $d=1.5$.}
\label{fig:qc2}
\end{center}
\end{figure}

\begin{figure}[htbp]
\begin{center}
\includegraphics[width=\columnwidth]{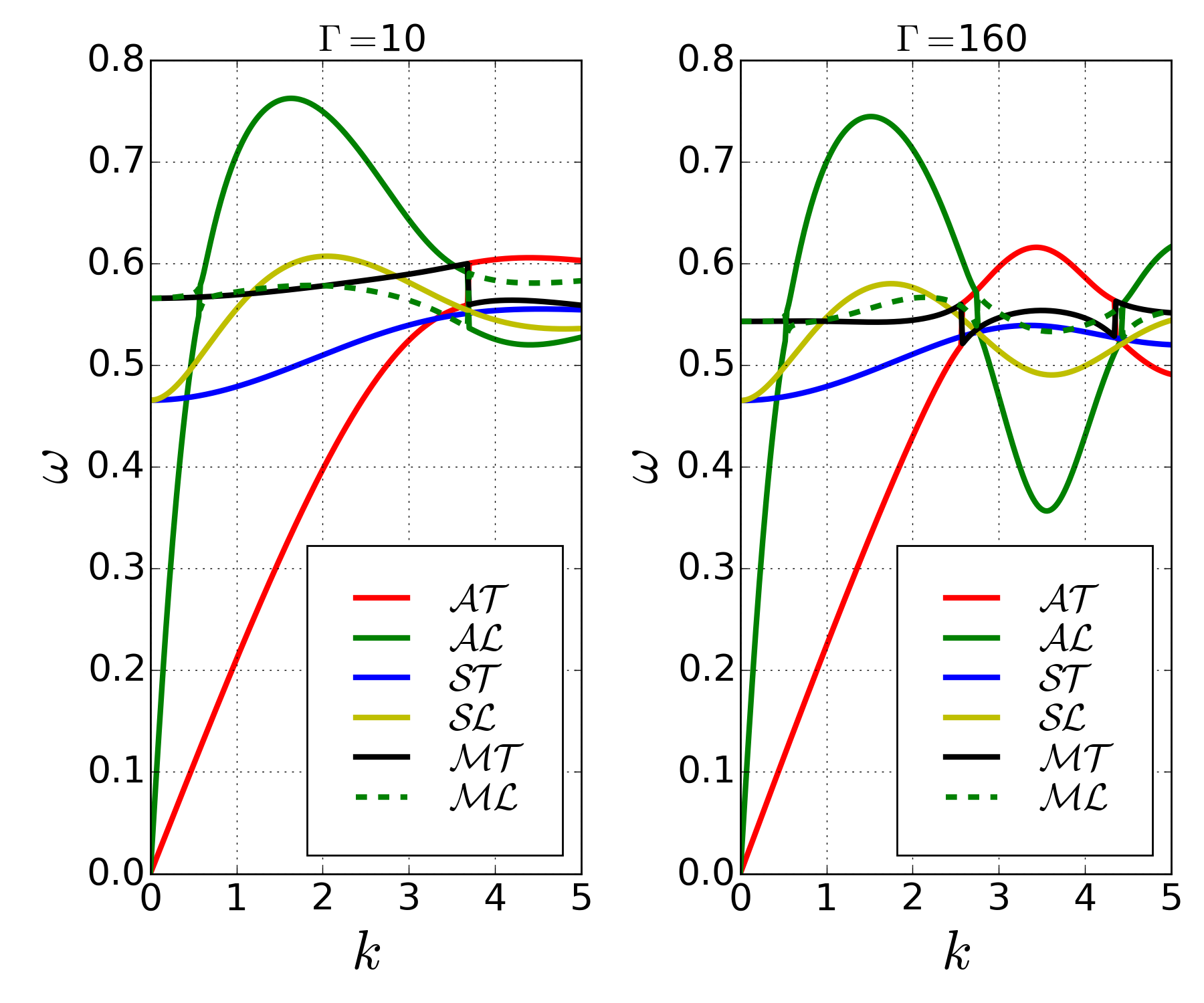}
\caption{QLCA dispersion curves of the six collective modes of the trilayer system, for moderate coupling $\Gamma=10$, and high coupling $\Gamma=160$ values, at $d=0.5$.}
\label{fig:qc3}
\end{center}
\end{figure}

\begin{figure}[htbp]
\begin{center}
\includegraphics[width=\columnwidth]{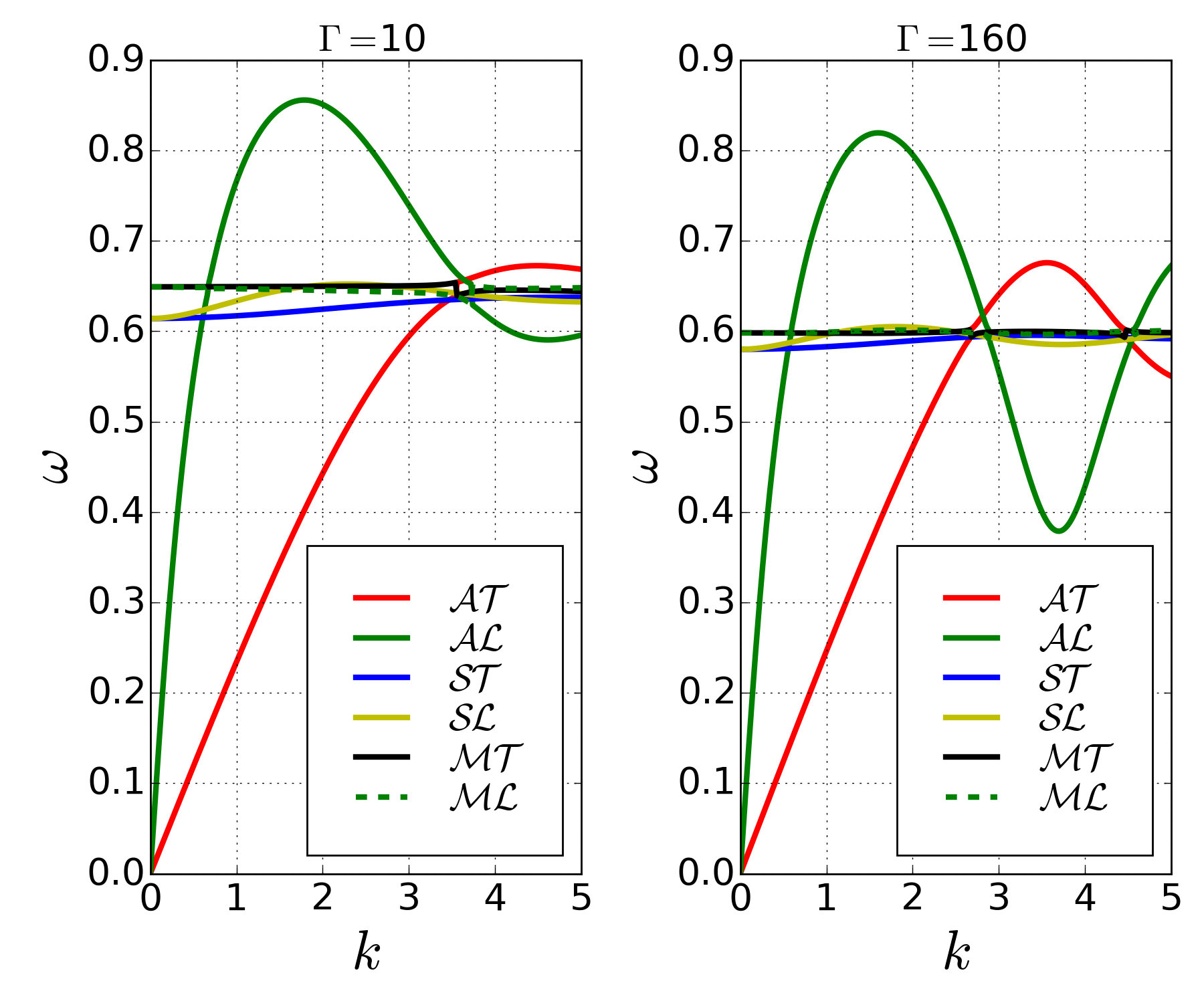}
\caption{QLCA dispersion curves of the six collective modes of the trilayer system, for moderate coupling $\Gamma=10$, and high coupling $\Gamma=160$ values, at $d=0.2$.}
\label{fig:qc4}
\end{center}
\end{figure}

\begin{figure}[htbp]
\begin{center}
\includegraphics[width=0.8\columnwidth]{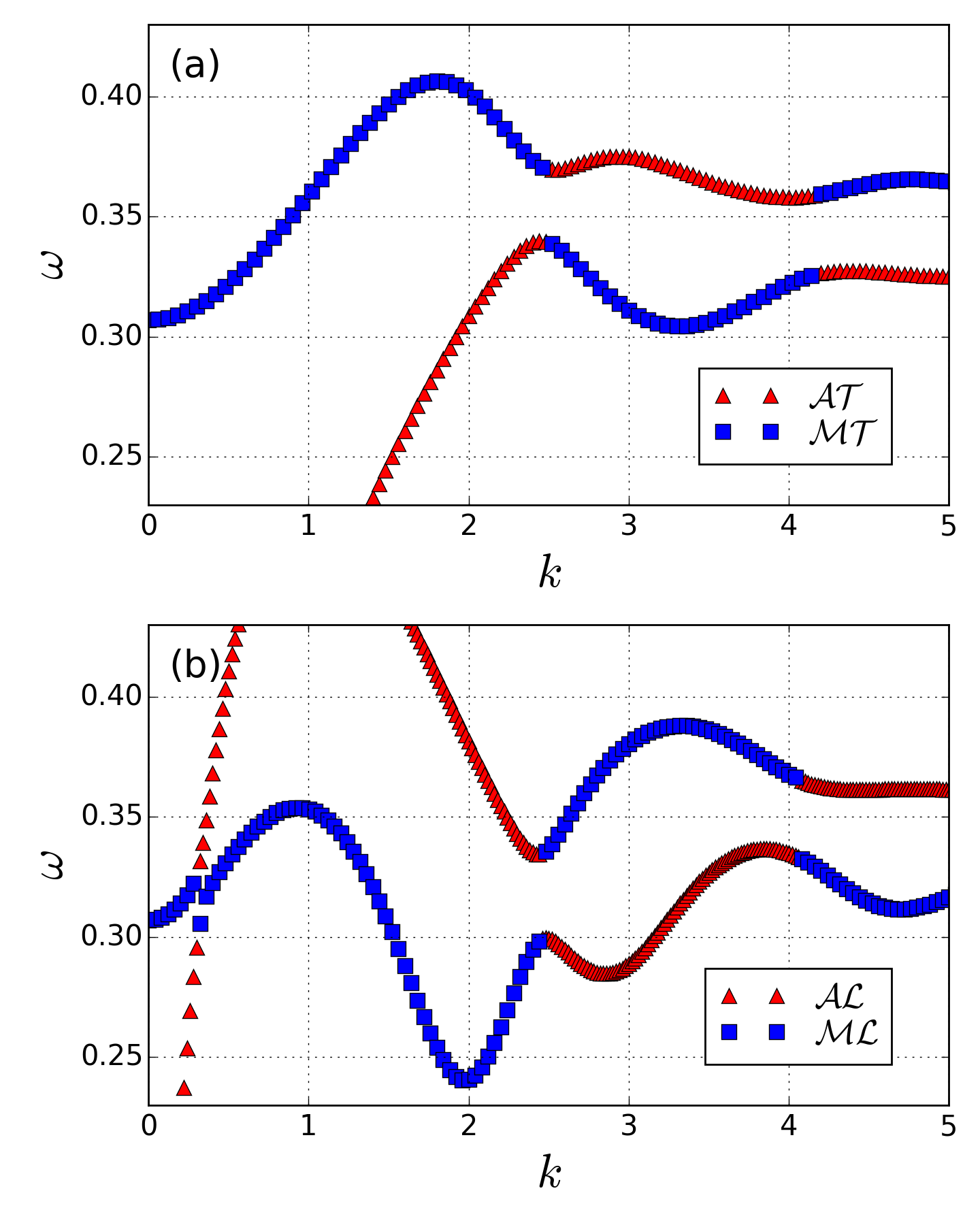}
\caption{Close-up of the QLCA dispersion of the $\mathcal{A}$ and $\mathcal{M}$ modes for $\Gamma$=160, at $d=1.5$. Note the avoided crossing and switch of polarizations between the modes.}
\label{fig:zma}
\end{center}
\end{figure}

\begin{figure}[htbp]
\begin{center}
\includegraphics[width=0.8\columnwidth]{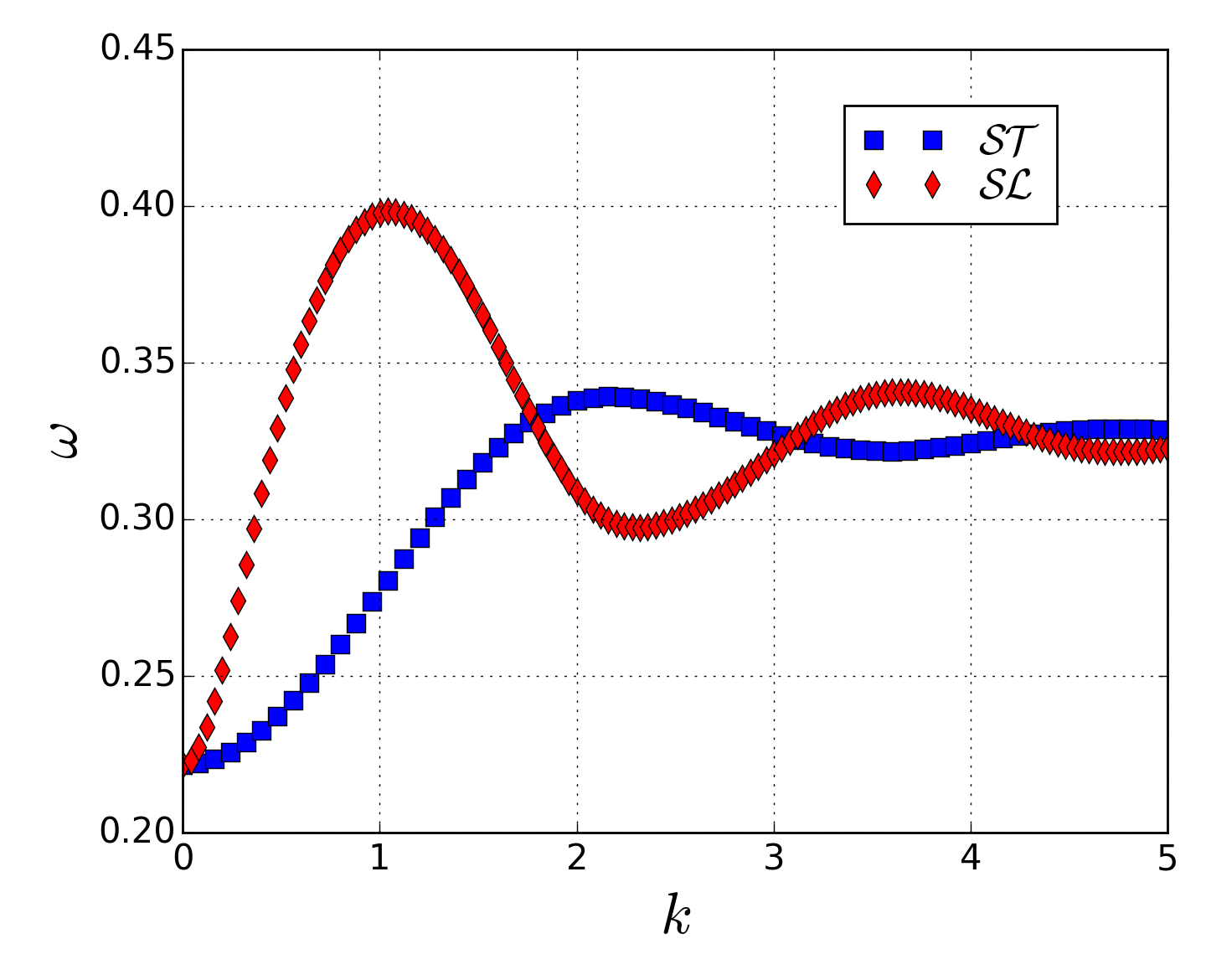}
\caption{Close-up of the QLCA dispersion of the $\mathcal{SL}$ and $\mathcal{ST}$ modes for $\Gamma$=160 at $d=1.5$. The simple crossing indicates that the two modes do not interact.}
\label{fig:zmb}
\end{center}
\end{figure}

The six calculated dispersion curves (DC), obtained with the aid of the input of the MD generated PDF-s are displayed in Figs.~\ref{fig:qc1} through \ref{fig:qc4} for moderate ($\Gamma=10$) and high ($\Gamma=160$) coupling values and for four different layer separations, $d=$~3.0, 1.5, 0.5, 0.2, corresponding to the three different equilibrium structures discussed in Section \ref{sec:structure}. They are represented by six continuous DC-s, portraying the six algebraic solutions. At the same time, the coloring of the DC-s follows the polarizations of the modes, which in general, are not continuous along these lines. More precisely, the polarization remains constant along the DC only if the mode in question is an eigenmode of a $k$-independently diagonalizable $2\times2$  sub-matrix.  This is the case for the Cartesian $\mathcal{L}$ and $\mathcal{T}$ modes and for the mode with $\mathcal{S}$ layer-space polarization. This behavior is illustrated in Figs. \ref{fig:zma} and Fig \ref{fig:zmb}
 
We note in passing that the physical reason for the $\mathcal{S}$ mode to decouple is that it represents the oscillations of a bilayer~\cite{Kalman1999}, composed of layers 2 and 3, in the presence of a dynamically inert layer 1.

In contrast to the above, the sub-matrix representing the $\mathcal{A}$ and $\mathcal{M}$ modes cannot $k$-independently be diagonalized and consequently these modes remain entangled with each other. The  root of the difference is that the $2\Leftrightarrow3$ symmetry that prevails in the $\mathcal{S}$ parent submatrix is broken in the $\mathcal{M}/\mathcal{A}$ submatrix. Moving along the respective DC-s, the mode polarizations change with $k$, up to the $k$ value where the two dispersion curves approach, but do not cross, each other. These are the so-called "avoided crossing" (AC) points, a well-known occurrence in many physical systems~\cite{Novotny2010,Baggioli2019}. Here, the AC behavior is mathematically  governed by the $Y$ matrix element, creating an AC point whenever $Y(k)=0$ occurs. Whether this may or may not happen depends on the system parameters, primarily on the layer separation $d$.

AC-like behavior is clearly visible in Figures 8 and 9. The details of  a set of AC-s are enlarged in Figs.~\ref{fig:zma} and \ref{fig:zmb}. In Fig.~\ref{fig:zma}(a) the transverse $\mathcal{A}$ and $\mathcal{M}$ DC-s approach each other at $k\approx 2.4$ and $k\approx 4.2$. These AC values are specific for the given $d$ and $\Gamma$, where the layer-space polarizations are exchanged and the curves show repulsive trajectories. As already pointed out, on the two sides of the AC points we identify the modes by the continuity of the polarizations of their eigenvectors, rather than by the continuity of the algebraic solutions. Equivalent phenomena, albeit at different $k$-values, can be observed for the longitudinal polarizations in Fig.~\ref{fig:zma}(b). A similar effect of interchanging eigenmodes in DC-s relating to other strongly coupled plasma systems has also been found in~\cite{Rosenberg2016}. 



The two limiting behaviors at $k=0$ and at $k\to\infty$ of the  DC-s are of interest.  At $k\to 0$ according to Eq.~(\ref{eq:fxyf}), the longitudinal and the transverse matrix elements become identical, which means that the longitudinal and transverse gaps become degenerate,  as they should, due to the isotropy of the system. The two gap frequencies at this point become
\begin{subequations} \label{eq:gapvs}
\begin{align}
\omega_\mathcal{S}^2(0)&=F(0)-D(0) \nonumber \\
&=-Y(0)-2D(0) \nonumber \\
&=\int_0^\infty [K_{12}(r)g_{12}(\rho)+2K_{23}(r)g_{23}(\rho)]\rho {\rm d} \rho \\
\omega_\mathcal{M}^2(0)&=-3Y(0) \nonumber \\ 
&=3\int_0^\infty K_{12}(rg_{12}(\rho)\rho {\rm d} \rho \\
K_{AB}(r)&=\frac{{\rm e}^{-\kappa r}}{3r^5}\left(\frac{1}{2}P\rho^2-Qr^2\right)
\end{align}
\end{subequations}
Thus the gap values depend on the interlayer correlation functions only. Remarkably, the $\mathcal{M}$-gap is fixed by the $g_{12}$ alone. In contrast, the $\mathcal{S}$-gap value in addition to the expected $g_{23}$, is affected by correlations with layer $1$ as well, even though this latter is dynamically inert. How close the two gap values are to each other is determined by the difference between $g_{12}$ and $g_{23}$.

At $k\to\infty$ the DC-s approach one of the  Einstein frequencies $\Omega_A$ (the oscillation frequency of a particle in layer $A$ in the presence of the frozen environment of all the other particles)
\begin{equation} \label{eq:einstein}
\Omega_A^2=\sum_{B=1,2,3}M_{xx}^{AB}(0) =\frac{1}{2}\sum_{B=1,2,3}M_{yy}^{AB}(0)
\end{equation}
which yields 
\begin{subequations} \label{eq:einst 1,2}
\begin{align}
\Omega_1^2= \int_0^\infty [&K_{11}(r)g_{11}(\rho)+2K_{12}(r)g_{12}(\rho)]\rho {\rm d} \rho \\
\Omega_2^2= \int_0^\infty [&K_{22}(r)g_{22}(\rho)+K_{12}(r)g_{12}(\rho) \nonumber \\
&+K_{23}(rg_{23}(\rho) ](\rho)\rho {\rm d} \rho 
\end{align}
\end{subequations}
A further case of interest is the $d\to 0$ limit, where- because of the substitutional disorder all the PDF-s have to become identical, as also verified  by the MD simulations: $g_{11}(\rho)=g_{22/33}(\rho)$ = $g_{12(\rho)}=g_{23}(\rho)$. (cf. Section~\ref{sec:structure}). In this case, the eigenvalues become  $(E-Y,E-Y,E+2Y)$. $E+2Y$ represents the acoustic ($\mathcal{A}$) mode, while $E-Y$ represents the $\mathcal{S}$ and $\mathcal{M}$ modes that degenerate into a single gapped mode. In view of the equality of the PDF-s and recalling Eq.~(\ref{eq:mat1}), we have
\begin{subequations}
\label{eq:eigenAB}
\begin{align}
\omega_{\mathcal{S},\mathcal{M}}^2(k)&=3M^{11}(0),    \\
\omega_\mathcal{A}^2(k)&=3[M^{11}(0)-M^{11}(k)].   
\end{align}      
\end{subequations}
Thus, the degenerate single  gapped mode is independent of $k$. Exploiting $3M^{11}=M^{\rm total}$, where $M^{\rm total}$ is the $M$ matrix calculated for a projected single layer, the gap frequency turns out to be identical to the Einstein frequency $\Omega_{total}$ of the projected single layer. As to the $\mathcal{AL}$ and $\mathcal{AT}$ modes, they become the corresponding acoustic modes of the projected single layer.


\section{\label{sec:MD} Molecular Dynamics simulation}

In the MD simulations the Newtonian equations of motion of each particle with mass $m$ and charge $q$ are integrated, restricted to in-plane motion only, applying the velocity-Verlet scheme using a time-step of $\Delta t =0.02/\omega_{\rm p}$. The inter-particle Yukawa forces are summed up for particle pairs with three-dimensional spatial separations smaller than a cut-off radius, chosen to be $R_{\rm cutoff} = 34 a$, taking into account the periodic boundary conditions. The $N=6000$ particles are first assigned in three equal parts to one of the layers with random initial in-plane positions in a square simulation cell. A velocity back-scaling ``thermostat'' is applied for the initial 20\,000 time-steps, after which the measurements are started without any thermostat for a period of 200\,000 time-steps. PDF and dynamical fluctuation data are recorded during this ``measurement'' phase.

From the MD simulations, the collective mode dispersion is determined by examining the current fluctuation spectra. Due to the structural isotropy of a liquid system, the modes may have only two, longitudinal or transverse, Cartesian polarizations. The periodicity implied by the boundary condition requires that the possible magnitudes of the wave vector $k$ are $k_n = n k_{\rm min}$, with $n$ being positive integers and $k_{\rm min}=2\pi/H$, where $H$ is the linear size of the simulation box.

The longitudinal and transverse current fluctuation spectra (current correlation functions) $L(k,\omega)$ and $T(k,\omega)$, are defined as
\begin{subequations}
\label{eq:cur7}
\begin{align}
   L_{AB}(k,\omega) &= \frac{2\pi}{\sqrt{N_A N_B}}\big\langle \widetilde{\lambda}_A(k,\omega) \widetilde{\lambda}_B(-k,-\omega)\big\rangle,  \\
   T_{AB}(k,\omega) &= \frac{2\pi}{\sqrt{N_A N_B}}\big\langle \widetilde{\tau}_A(k,\omega) \widetilde{\tau}_B(-k,-\omega)\big\rangle,
\end{align}
\end{subequations}
where the $A, B$ indices represent the layers; $N_A$, $N_B$ are the particle numbers of each layer, in our case $N_A = N_B$. $\widetilde{\lambda}_A(k,\omega)$ and $\widetilde{\tau}_A(k,\omega)$ are the Fourier transforms of $\lambda_A(k,t)$ and $\tau_A(k,t)$.  In order to improve the signal-to-noise ratio of the data we average multiple spectra obtained from subsequent time series for the microscopic quantities, which is represented by the $\langle...\rangle$ notation in Eq.~\ref{eq:cur7}. The microscopic currents are   
\begin{subequations}
\label{eq:cur6}
\begin{align}
   \lambda_A(k,t) &= \sum_{j=1}^{N_A} v_{x,j} \exp\left(i k_x x_j\right),  \\ 
   \tau_A(k,t)    &= \sum_{j=1}^{N_A} v_{y,j} \exp\left(i k_x x_j\right),
\end{align}
\end{subequations}
where the subscript $j$ labels the particles, the summation runs over all particles in layer $A$. 
    
$L(k,\omega)$ and $T(k,\omega)$ obey the reality condition that requires, e.g., $L(k,\omega)=L^*(-k,-\omega)$, where * represents complex conjugate. Thus $L_{AA}(k,\omega)$ and $T_{AA}(k,\omega)$ are real valued, but $L_{AB}(k,\omega)$ and $T_{AB}(k,\omega)$, in general, may be complex. However, in a liquid, where the system obeys inversion symmetry $L_{AB}(k,\omega)$ and $T_{AB}(k,\omega)$ are also real. 

The dispersion relation for the modes is determined by identifying the peaks in the fluctuation spectra.


\section{\label{sec:MDresults}Collective mode spectrum: MD simulations}

The six modes predicted by the QLCA and discussed in Section~\ref{sec:QLCA} can be identified from the analysis of the eight distinct longitudinal and transverse  current fluctuation spectra $L_{AB}(k,\omega)$ and $T_{AB}(k,\omega)$, with  $AB = 11, 22, 12, 23$. While the longitudinal and transverse spatial polarizations appear automatically separated in their respective $L$ and $T$ fluctuation spectra, in general a mixture of the three $\mathcal{A}$, $\mathcal{M}$, and $\mathcal{S}$ modes with different layer-space polarizations shows up in each of the $11, 22, 12, 23$ fluctuation spectra. The $11, 22$ peak signatures are necessarily positive, but the $12, 23$ peaks may have positive or negative signature, depending on whether in that particular mode the two layers involved oscillate in-phase or out-of-phase. 

\begin{figure}[htbp]
\begin{center}
\includegraphics[width=\columnwidth]{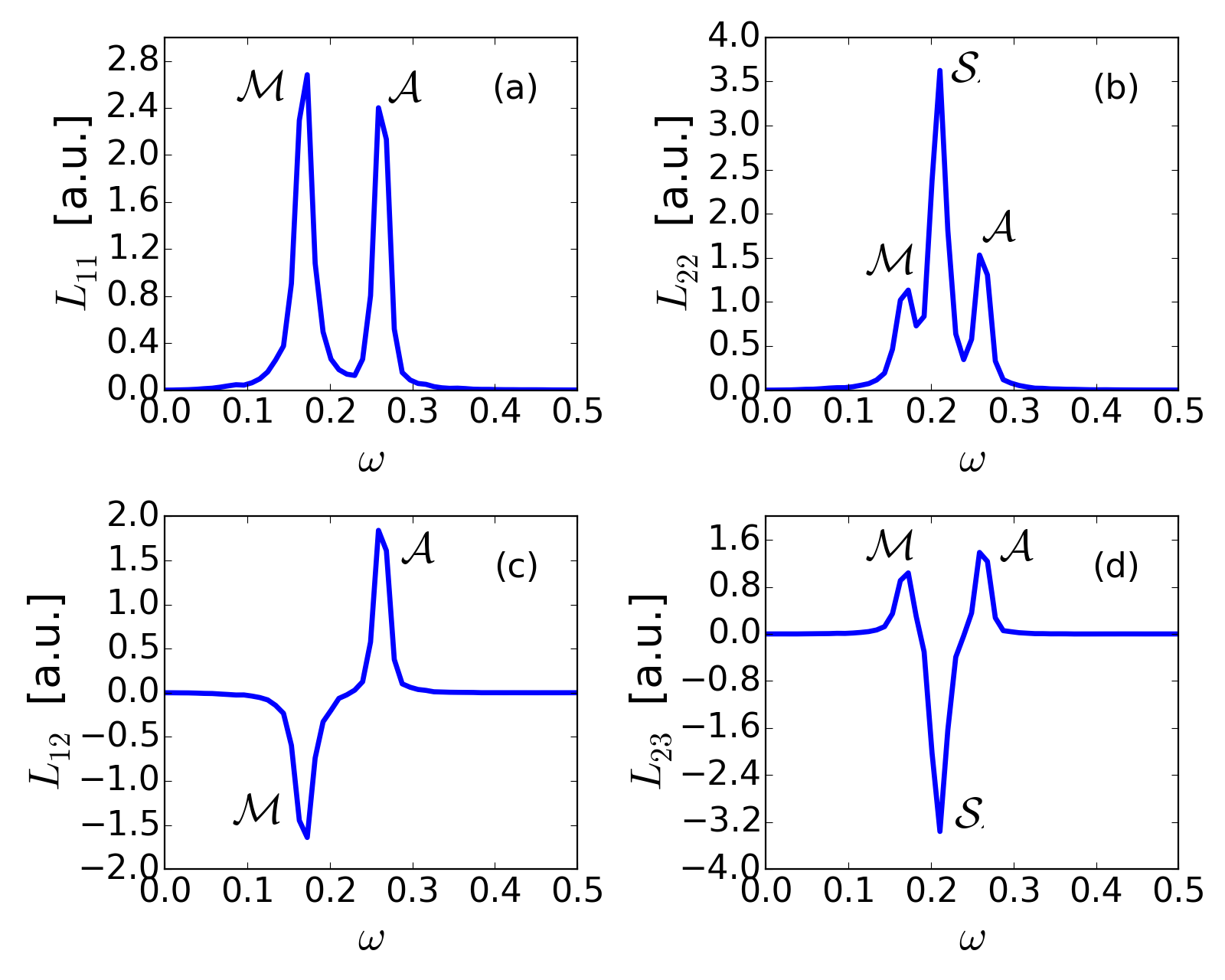}
\caption{Longitudinal current fluctuation spectra at $k=0.32$ for $\Gamma=160$, at $d=3.0$, (a) $L_{11}$, (b) $L_{22}$, (c) $L_{12}$, (d) $L_{23}$. The labeling of the peaks identifies the corresponding mode. }
\label{fig:Lalls}
\end{center}
\end{figure}

These structures are illustrated in Fig.~\ref{fig:Lalls} that shows MD simulation results for the $L_{AB}(k,\omega)$ spectra, for $k=0.32$, $\Gamma$=160, at  $d=3.0$. At this $k$ value the positions (frequencies) of the peaks are well separated. A similar picture emerges for the $T(k,\omega)$ correlation functions.

We observe that in each of the computed spectra ($L_{AB}$ and $T_{AB}$) the contribution of the principal oscillation modes ($\mathcal{A}$, $\mathcal{M}$, $\mathcal{S}$) appear mixed. As example, $L_{11}$ contains two peaks, one originating from the $\mathcal{AL}$ and one from the $\mathcal{ML}$ modes both with positive signs, while the same modes contributing to $L_{12}$ show respectively  positive (for $\mathcal{AL}$) and negative (for $\mathcal{ML}$) peaks. Making use of this parity property one can apply simple linear combinations in order to separate out the contribution of a selected mode:
\begin{eqnarray*}     
        L_{11}-L_{12} & \rightarrow & ~\mathcal{ML} ~ {\rm mode}, \nonumber \\
        L_{11}+L_{12} & \rightarrow & ~\mathcal{AL} ~ {\rm mode}, \nonumber \\
        L_{22}-L_{23} & \rightarrow & ~\mathcal{SL} ~ {\rm mode} \nonumber,       
\end{eqnarray*}
The last relationship, in fact, represents according to Eq.~\ref{eq:MMM} the exact diagonalization that fully decouples the $\mathcal{S}$ mode \footnote{This part is authorized by GK, PH, ZD.}. The others are heuristic constructions that almost fully cancel the presence of the other modes in the spectrum. A supplementary method that we also apply in certain situations consists of creating a linear combination of all the four current fluctuation spectra with coefficients whose values are determined by optimization to achieve the best separation of modes.

\begin{figure}[htbp]
\begin{center}
\includegraphics[width=\columnwidth]{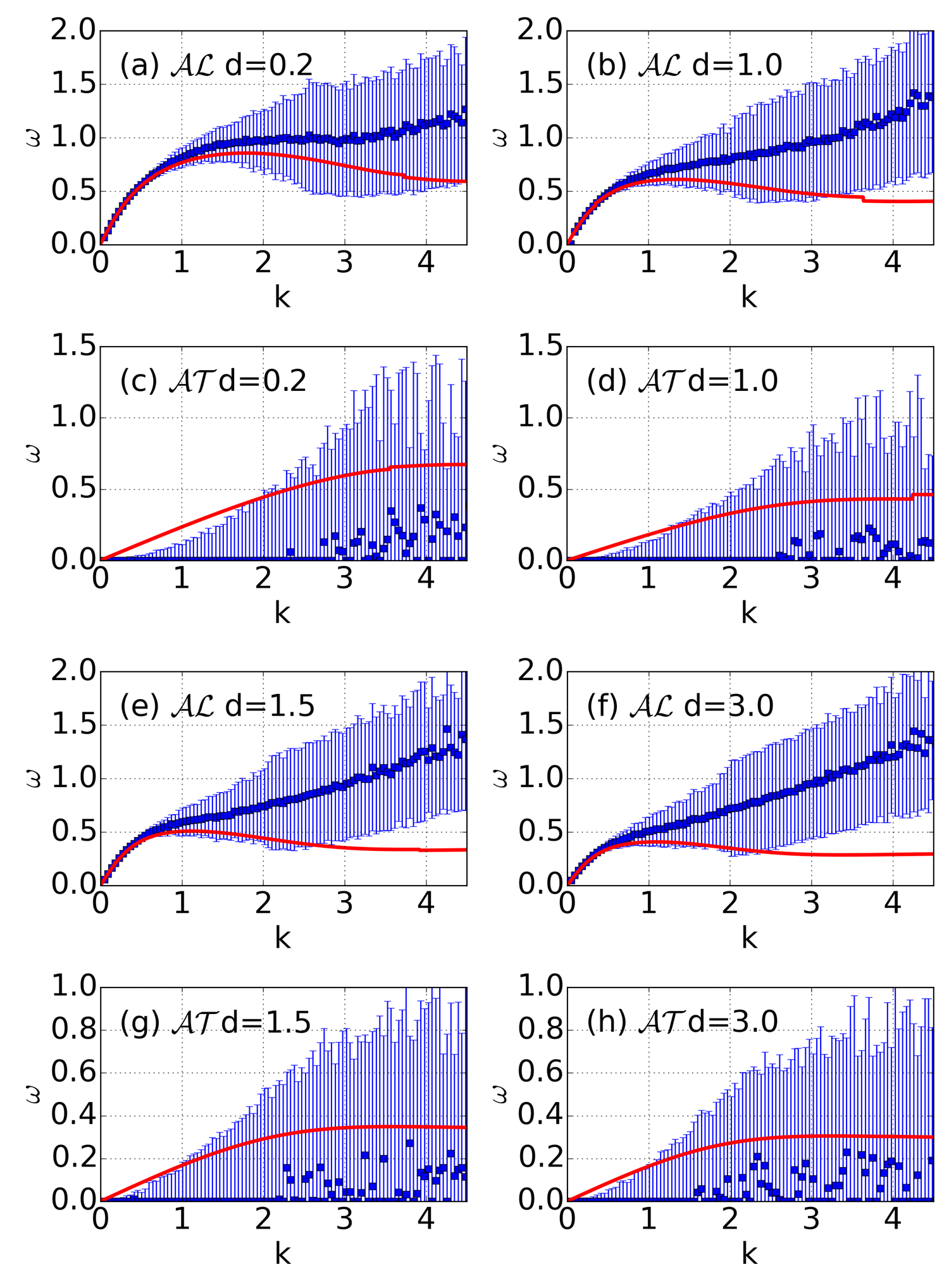}
\caption{Collective mode dispersion  comparison between MD (blue squares) and QLCA (red lines) for the $\mathcal{A}$ mode, for $\Gamma$=10 at different $d$ values in the range $0.2\leq{d}\leq3.0$. The vertical bars indicate the Gaussian width of the spectral peaks, as explained in the text. }
\label{fig:mq1A}
\end{center}
\end{figure}

\begin{figure}[htbp]
\begin{center}
\includegraphics[width=\columnwidth]{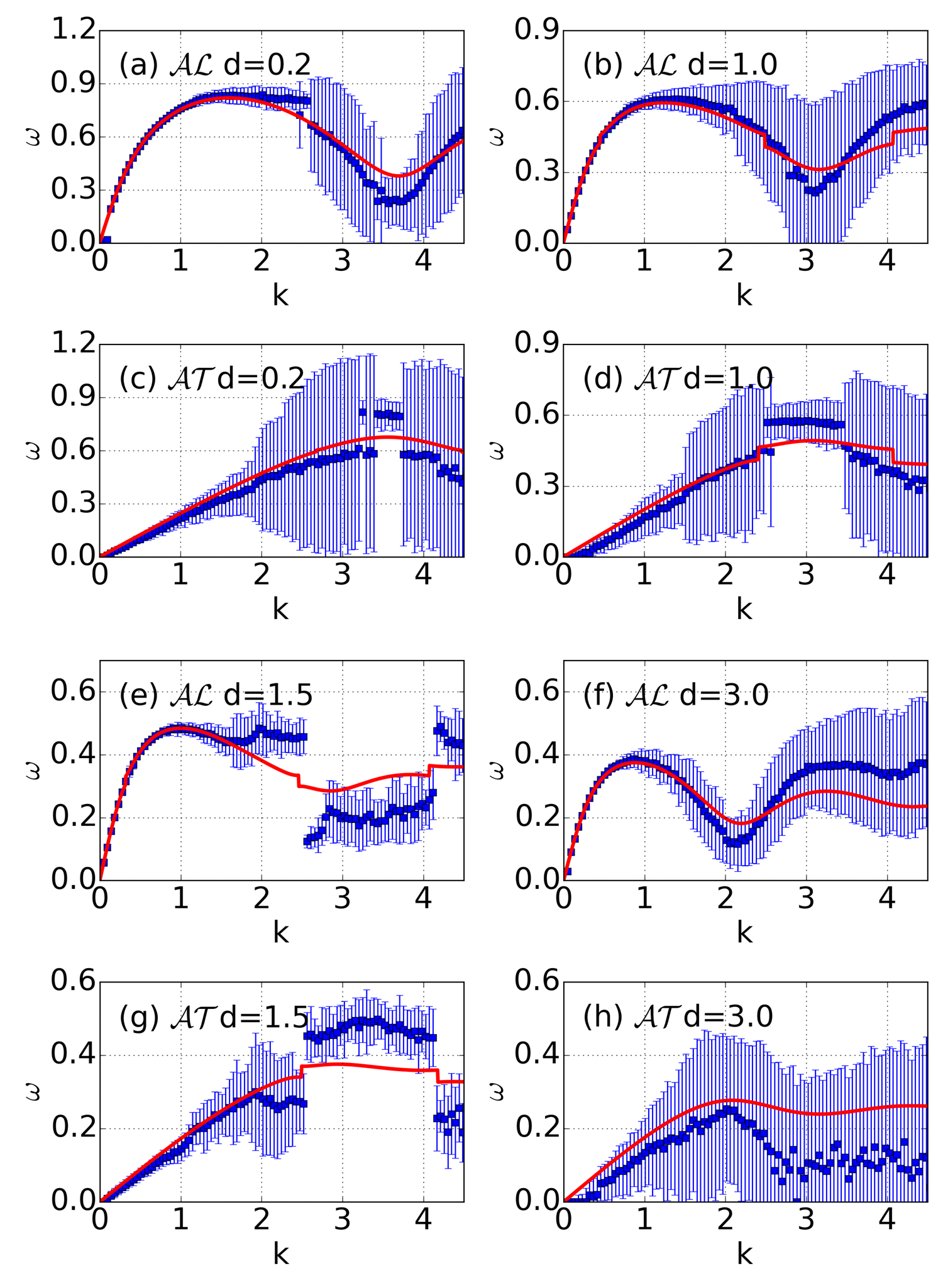}
\caption{The same as Fig.~\ref{fig:mq1A} for $\Gamma=160$. Compare the avoided crossing points in panels (e) and (g) at $k=2.6$ with their counterparts in panels (a) and (c), Fig.~\ref{fig:mq4}. See also Fig.~\ref{fig:zmb} for the details of the predicted QLCA behavior. Other apparent AC points at lower $d$ values lack simple  interpretation, as explained in the text.}
\label{fig:mq1B}
\end{center}
\end{figure}

\begin{figure}[htbp]
\begin{center}
\includegraphics[width=\columnwidth]{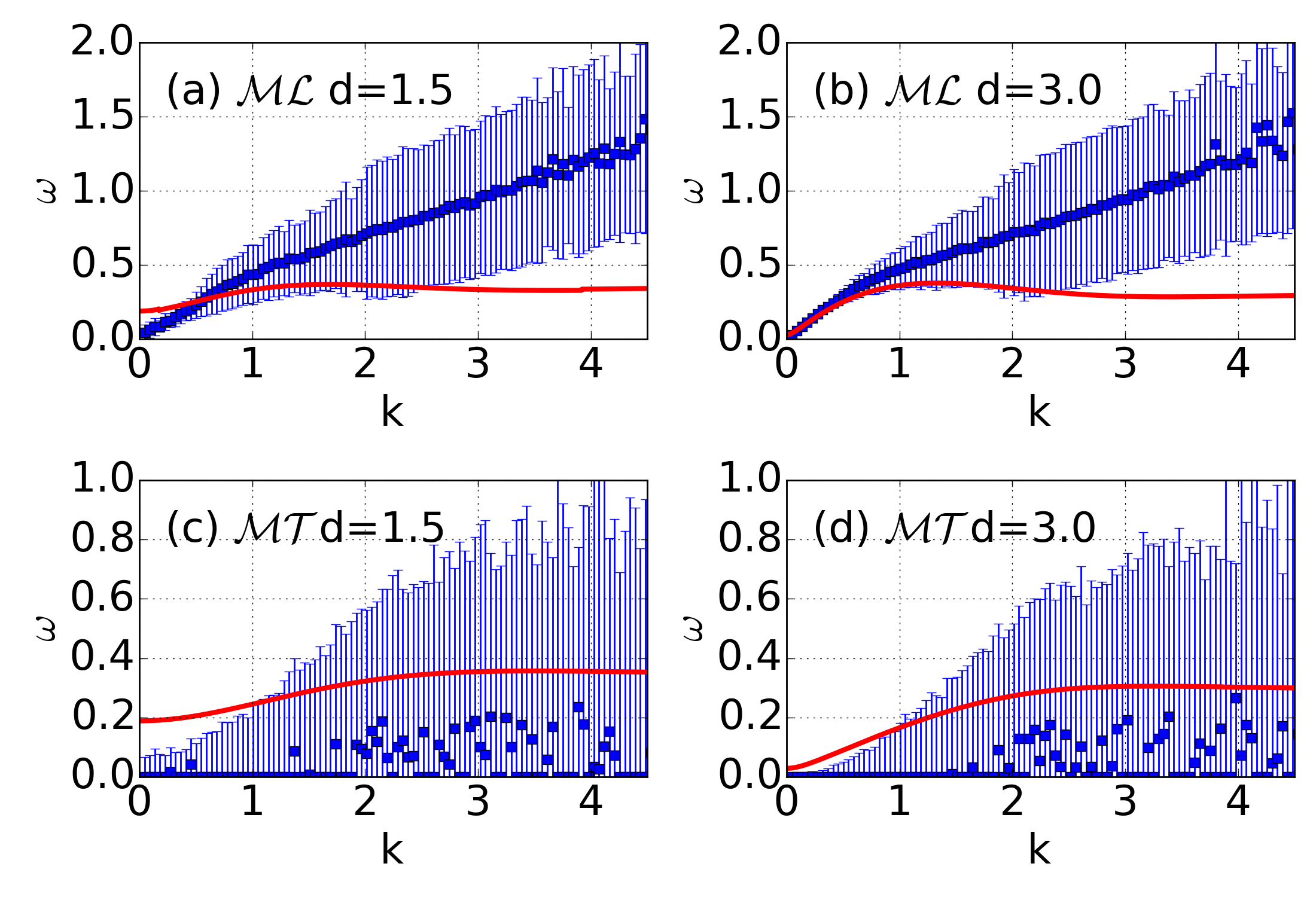}
\caption{The same as Fig.~\ref{fig:mq1A} for the $\mathcal{M}$ mode, for $\Gamma$=10, at $d=1.5$ and $d=3.0$ values. $d<1.5$ layer separations are not shown, for reasons explained in the text. }
\label{fig:mq3}
\end{center}
\end{figure}

\begin{figure}[htbp]
\begin{center}
\includegraphics[width=\columnwidth]{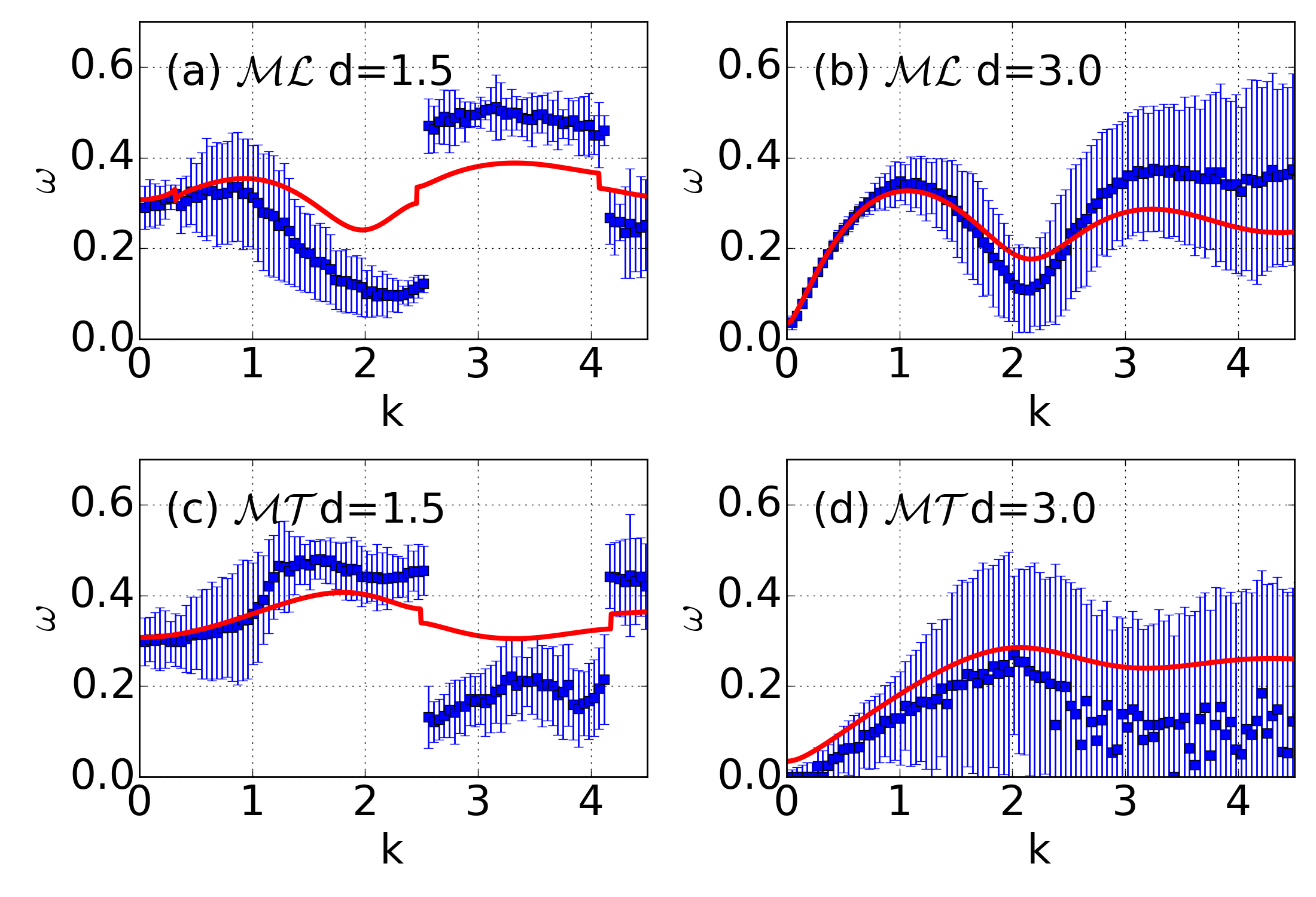}
\caption{The same as Fig.~\ref{fig:mq3} for $\Gamma$=160. Compare the avoided crossing points in panels (a) and c) at $k=2.6$ with their  counterparts in panels (e) and (g), Fig.~\ref{fig:mq1B} See also Fig.~\ref{fig:zmb} for the details of the predicted QLCA behavior. 
}
\label{fig:mq4}
\end{center}
\end{figure}

\begin{figure}[htbp]
\begin{center}
\includegraphics[width=\columnwidth]{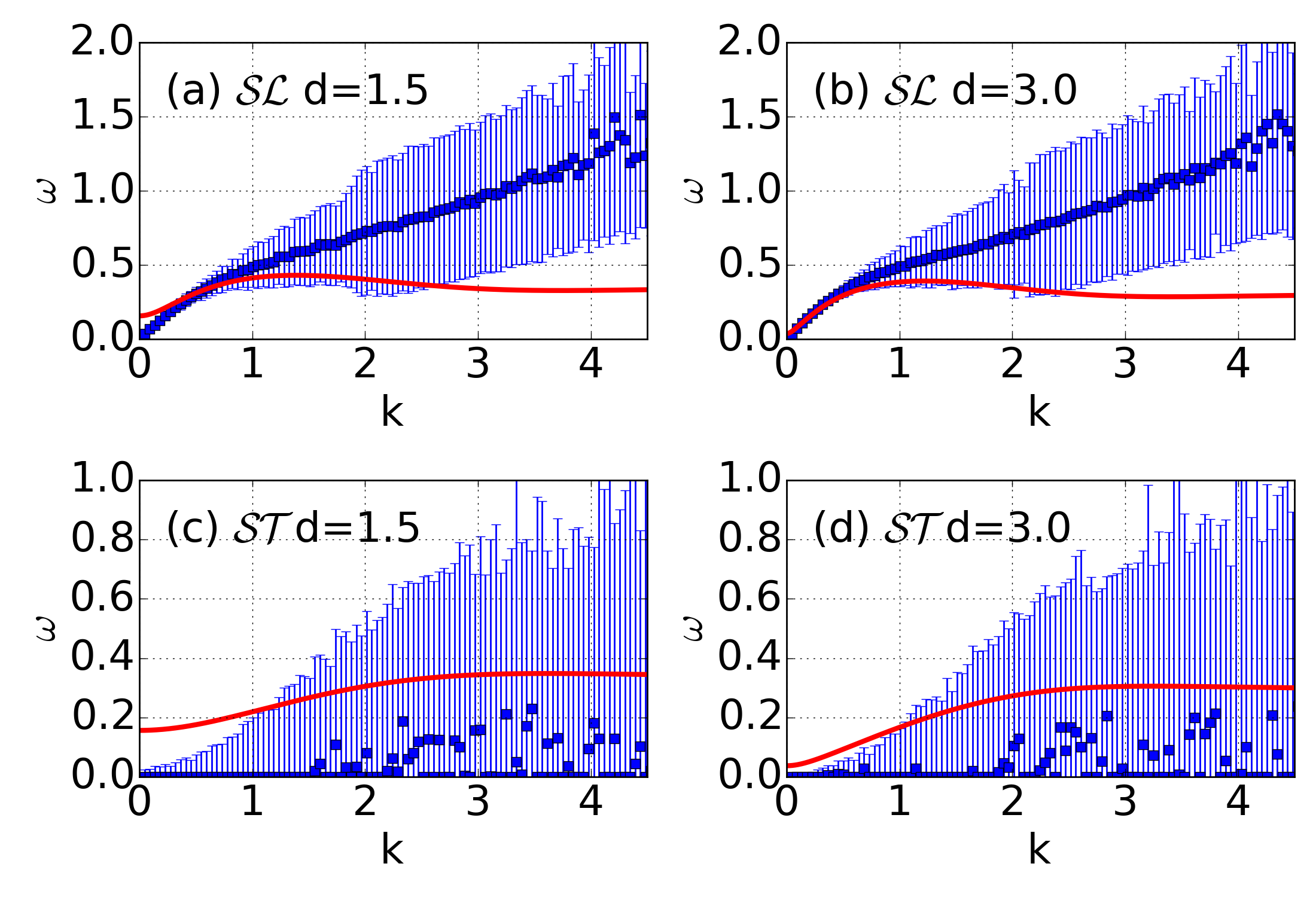}
\caption{The same as Fig.~\ref{fig:mq3} for the $\mathcal{S}$ mode, for $\Gamma$=10.}
\label{fig:mq5}
\end{center}
\end{figure}

\begin{figure}[htbp]
\begin{center}
\includegraphics[width=\columnwidth]{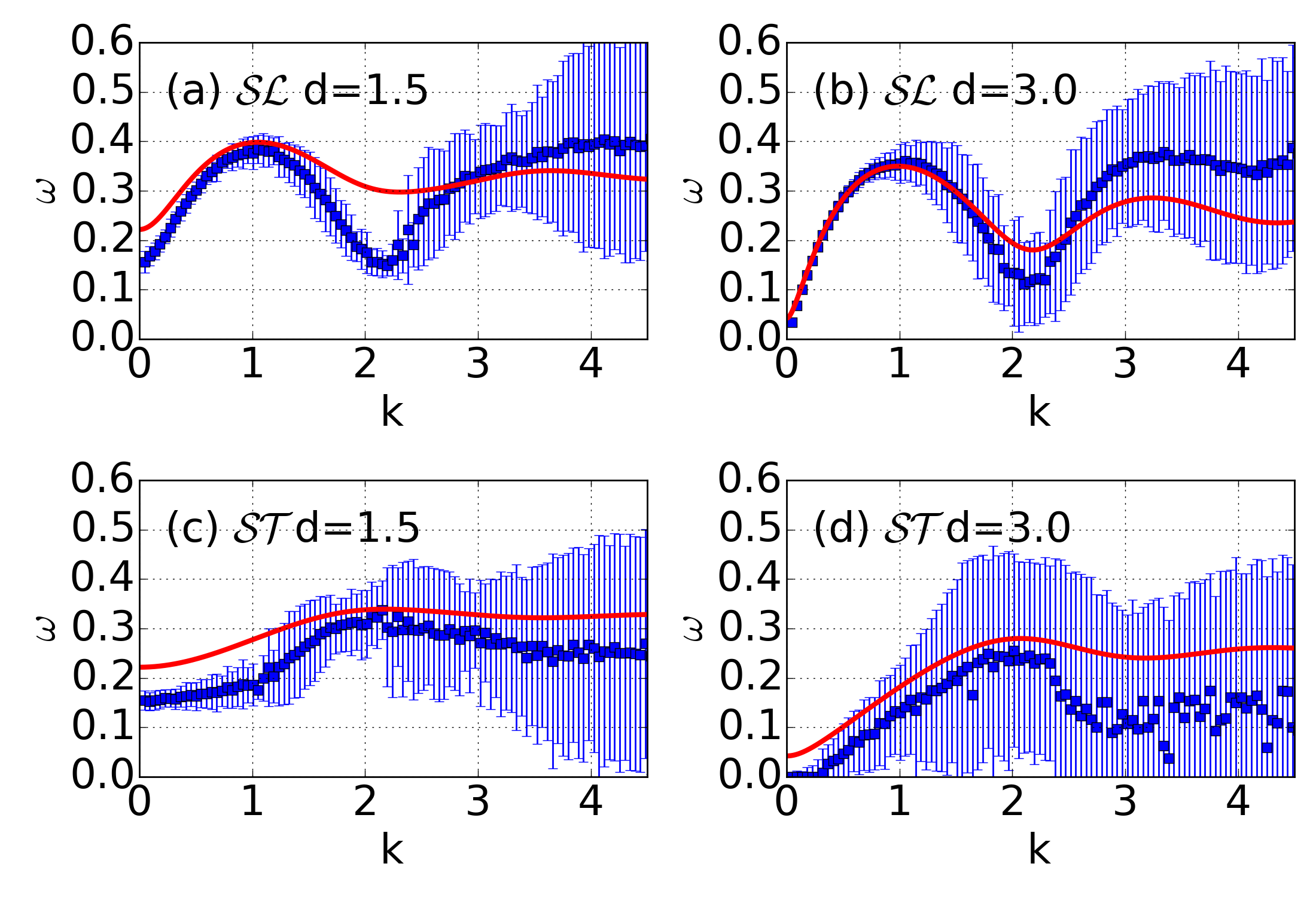}
\caption{The same as Fig.~\ref{fig:mq3} for the $\mathcal{S}$ mode, for $\Gamma$=160. }
\label{fig:mq6}
\end{center}
\end{figure}

Dispersion curves obtained from the MD simulations for each of the collective modes, together with their corresponding QLCA counterparts are displayed in Figs.~\ref{fig:mq3} through \ref{fig:mq6}. The organization of these Figures is as follows. Each Figure depicts the pair of the $\mathcal{L}$ and the $\mathcal{T}$ spatial polarizations of one of the three basic $\mathcal{A}$, $\mathcal{M}$ and $\mathcal{S}$ modes, for two values of the layer separation, $d=1.5$ and $d=3.0$. Additionally, in Fig.~\ref{fig:mq1A} and Fig.~\ref{fig:mq1B} we display for $d=0.2$ and $d=1.0$ the dispersions of the in-phase $\mathcal{AL}$ and $\mathcal{AT}$ modes. Dispersion curves for the gapped modes for lower than $d=1.5$ values are not shown for reasons explained below. The blue square symbols correspond to the peak positions of the MD generated $L$ and $T$ current fluctuation spectra, with vertical bars showing the full width at half maximum based on a Gaussian fit to the spectral peaks. The red lines represent the QLCA dispersion curves. The plots are given for moderate $\Gamma=10$ and high $\Gamma=160$ coupling values.

In assessing the reliability of the QLCA we observe that the agreement between the MD and QLCA results is much more pronounced at $\Gamma=160$ than at $\Gamma=10$. This is expected, since the QLCA is a strong coupling approximation. There is also a difference in the quality of match between the behavior of the acoustic and gapped modes. Focusing on the acoustic modes, for the longitudinal polarization there is an overall good, for low $k$ values excellent, agreement between the MD and QLCA results in all cases. For the transverse polarization, good agreement is obtained only for high $\Gamma$, because in a weakly coupled liquid no transverse (shear) excitation can exist. Even at strong coupling, the $\mathcal{T}$ mode does not extend down to $k=0$, rather cuts off at a small, but finite $k$ value, as also noted in several earlier studies~\cite{Kalman2000,Ohta2000,Khrapak2015,Mithen2011,Murillo2000,Nosenko2006}. With increasing $k$ values, the error bars grow longer and thermal effects make the dispersion deviate, especially in the weaker coupling case, from the QLCA prediction. 

As to the gapped modes, they can be identified in the MD data only in the strong coupling case. There the agreement is between good and fair, better for the high frequency $\mathcal{M}$ than the lower frequency $\mathcal{S}$ modes.

For high $\Gamma$, the MD simulations also show the existence of some anomalous branches, in the spectrum, not predicted by the QLCA. For the sake of clarity, these modes are not displayed in the subsequent Figures, but are discussed separately below.

The roton minimum that was predicted in~\cite{Kalman2010} to occur in the dispersion at higher $k$ values in all strongly coupled plasma systems is indeed visible in all graphs at high $\Gamma$, although the QLCA has the tendency to under-represent the oscillation amplitudes of the dispersion curves: this feature has also been noted earlier~\cite{Golden2010}. 

In some of the dispersion curves discontinuous jumps from one branch to a neighboring one can be observed (It should be kept in mind that the modes are labelled by their polarizations). In Figs.~\ref{fig:mq4}c and \ref{fig:mq6}c, e. g., this behavior is the clear verification \footnote{This part has been authored by GJK, PH and ZD.} of the AC phenomenon --~a remarkable  feature of the trilayer~-- taking place between two QLCA modes (cf. Fig.~\ref{fig:qc2}b). In other cases non-QLCA excitations are also involved (see below and Section~\ref{sec:envelope}), whose details are of little interest here (cf. Fig.~\ref{fig:mq1B}, panels (a) through (d)).

\begin{figure}[htbp]
  \centering
  \includegraphics[width=\columnwidth]{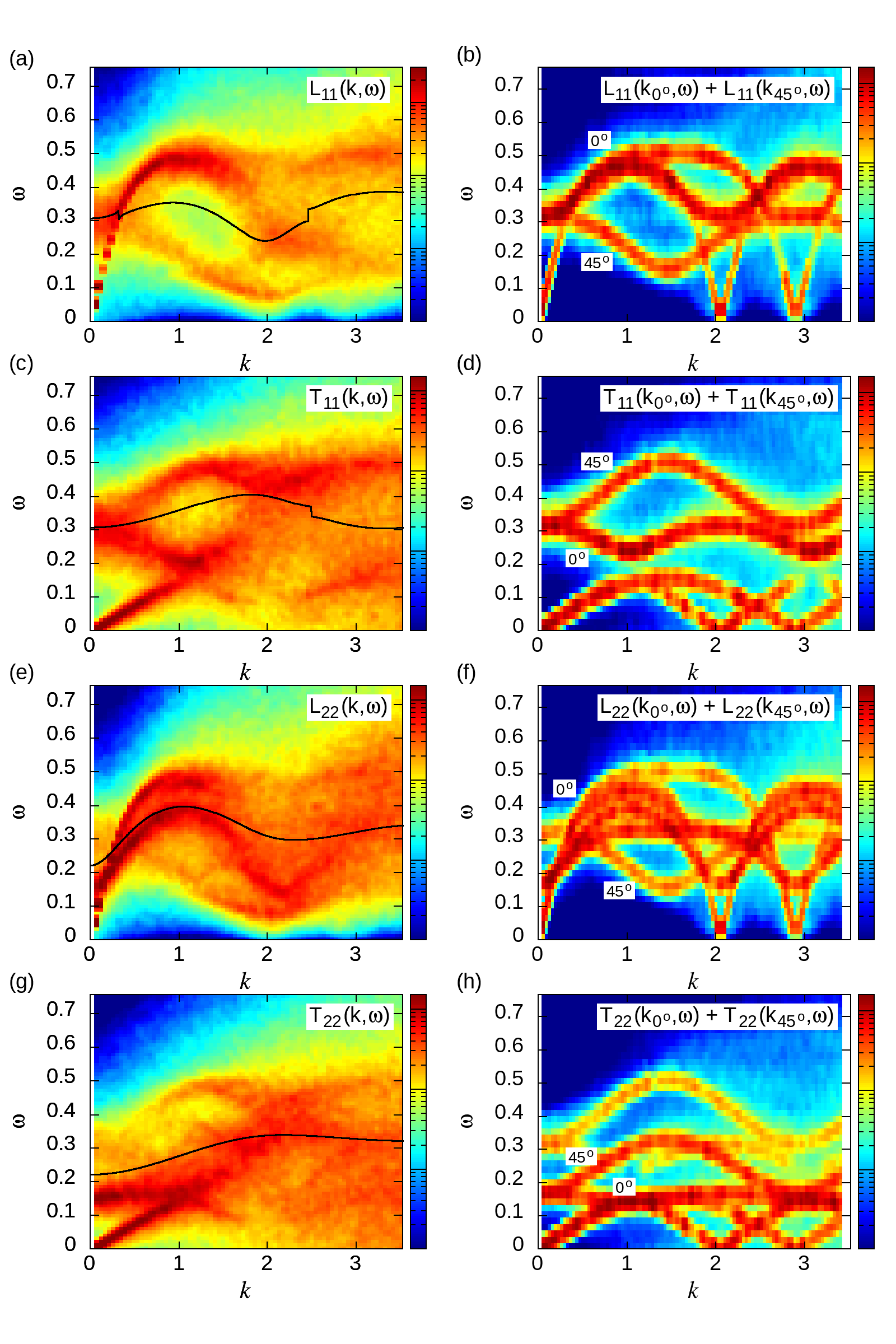}
  \caption{Comparison of current fluctuation spectra between the  liquid for $\Gamma=160$, (left column) and the lattice for $\Gamma=1000$, (right column) phases at $d=1.5$. Displayed are the (a,b): $L_{11}(k,\omega)$, (c,d): $T_{11}(k,\omega)$, (e,f): $L_{22}(k,\omega)$, (g,h): $T_{22}(k,\omega)$  spectra. In the liquid spectra black lines show the corresponding QLCA mode dispersions:  $\mathcal{ML}$ in (a), $\mathcal{MT}$ in (c), $\mathcal{SL}$ in (e), $\mathcal{ST}$ in (g). The MD simulations  are for the OS lattice system and for propagation directions $\phi=$~\ang{0} and $\phi=$~\ang{45}, as indicated for the corresponding  of the spectral features. The amplitudes are given in arbitrary units, therefore numerical values at the color bars are omitted. The color scale is logarithmic and covers three orders of magnitude.}
  \label{fig:summSpec}
\end{figure}

As noted above, additionally to the $2 \times 3$ modes predicted by the QLCA, an unexpected feature is revealed by the MD simulations. It is the appearance for $\Gamma>50$ of two optical branches, one with $L$, one with $T$ polarization. They are illustrated in Fig.~\ref{fig:summSpec}, where in contrast to the predicted two modes we see the appearance of three branches for each polarization. The two anomalous branches seem to be satellite excitations to the $\mathcal{ML}$ and $\mathcal{MT}$ modes, respectively (and designated as $\mathcal{MLX}$ and $\mathcal{MTX}$ branches). They share the $k=0$ gap frequency and the layer polarizations with their parent $\mathcal{ML}$ and $\mathcal{MT}$ modes. What distinguishes them though, is the slope of their dispersion curves at $k \to 0$: it is downward, while the parent modes' dispersion curves always have an upward slope.    

We believe that the source of this phenomenon can be found in the development of microcystals in the strongly coupled high $\Gamma$ liquid. The collective modes supported by the crystal lattice are identical to those appearing in the liquid, but for the fact that the lattice is anisotropic, while the liquid is not. As a result, the propagation characteristics of the modes depend on the angle $\phi$ between the $\vec {k}$ and a chosen principal axis of the lattice. Since the orientation of the microcrystals should be random, the QLCA calculated $\mathcal{ML}$ and $\mathcal{MT}$ modes for the microcrystals may be regarded as, in a certain sense, an angular average over $\phi$~\cite{Golden2002}. Such an average results in an upward slope, because this very same feature is exhibited by the dispersion curves in an overwhelmingly large angular domain of the unit cell~\cite{Hongthesis}. On the other hand, what we see as the $\mathcal{MLX}$ and $\mathcal {MTX}$ satellites can be understood as follows. There are propagation angles along which either of the lattice equivalents of the $\mathcal{ML}$ and $\mathcal{MT}$ modes develops a dispersion substantially different from the QLCA generated average. This is shown in Figs.~\ref{fig:summSpec}b and \ref{fig:summSpec}d  calculated for an OS crystal lattices structure at $\phi=0^\circ$ and $\phi=45^\circ$. It is the imprints of the $L$ and $T$ projections of these anomalous dispersions that appear as the $\mathcal{MLX}$ and $\mathcal{MTX}$ satellites, respectively, as illustrated in Figs.~\ref{fig:summSpec}a and \ref{fig:summSpec}c. We note that no satellites develop for the $\mathcal{S}$ modes, as may be seen in Figs.~\ref{fig:summSpec}e and \ref{fig:summSpec}g,  because their lattice equivalents never qualitatively deviate from their QLCA value (Figs.~\ref{fig:summSpec}f and \ref{fig:summSpec}h).

Our discussion in this Section has covered the behavior of the gapped modes in the domain $d>1.0$. When $d$ drops below the $d=1.0$ value, their good agreement with the QLCA ceases. What seems to happen is that the peaks belonging to the two gapped modes in the $L$ and $T$ fluctuation spectra weaken and when $d=1.0$ is reached they virtually disappear. At the same time, a new broad double-peaked feature appears in the spectrum, extending way beyond the QLCA predicted gap values. We will refer to this novel structure as the ``envelope'': its properties will be discussed in the next Section.

\section{Envelope formation} \label{sec:envelope}

\begin{figure}[htbp]
\begin{center}
\includegraphics[width=\columnwidth]{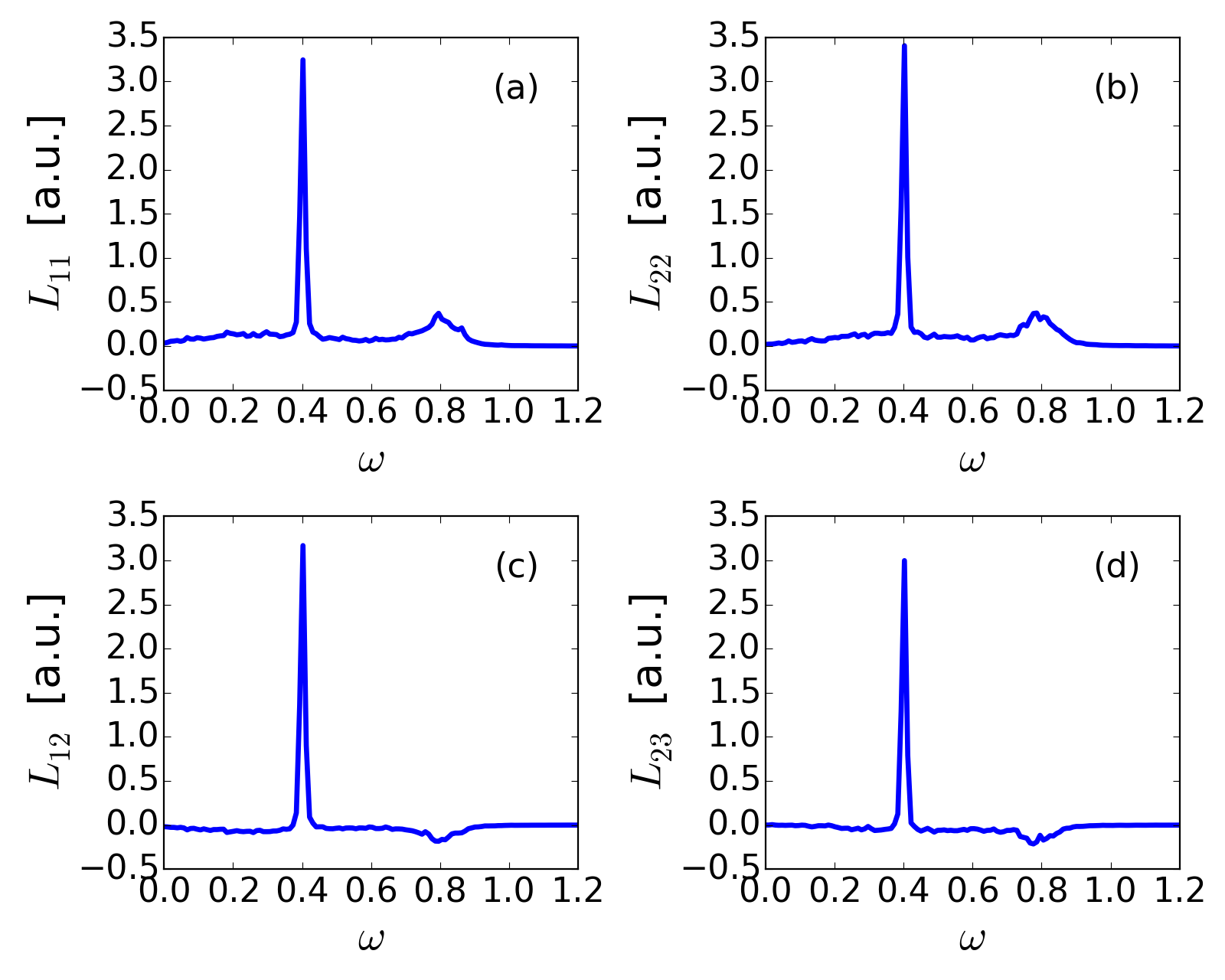}
\caption{Longitudinal current fluctuation spectra  $L_{11}$, $L_{22}$, $L_{12}$, and $L_{23}$ at a small layer separation, $d = 0.2$, $k = 0.32$ for $\Gamma=160$. Note the different patterns compared to the large $d$ case shown in Figure~\ref{fig:Lalls}. }
\label{fig:Lsd02}
\end{center}
\end{figure}

\begin{figure}[htbp]
\begin{center}
\includegraphics[width=\columnwidth]{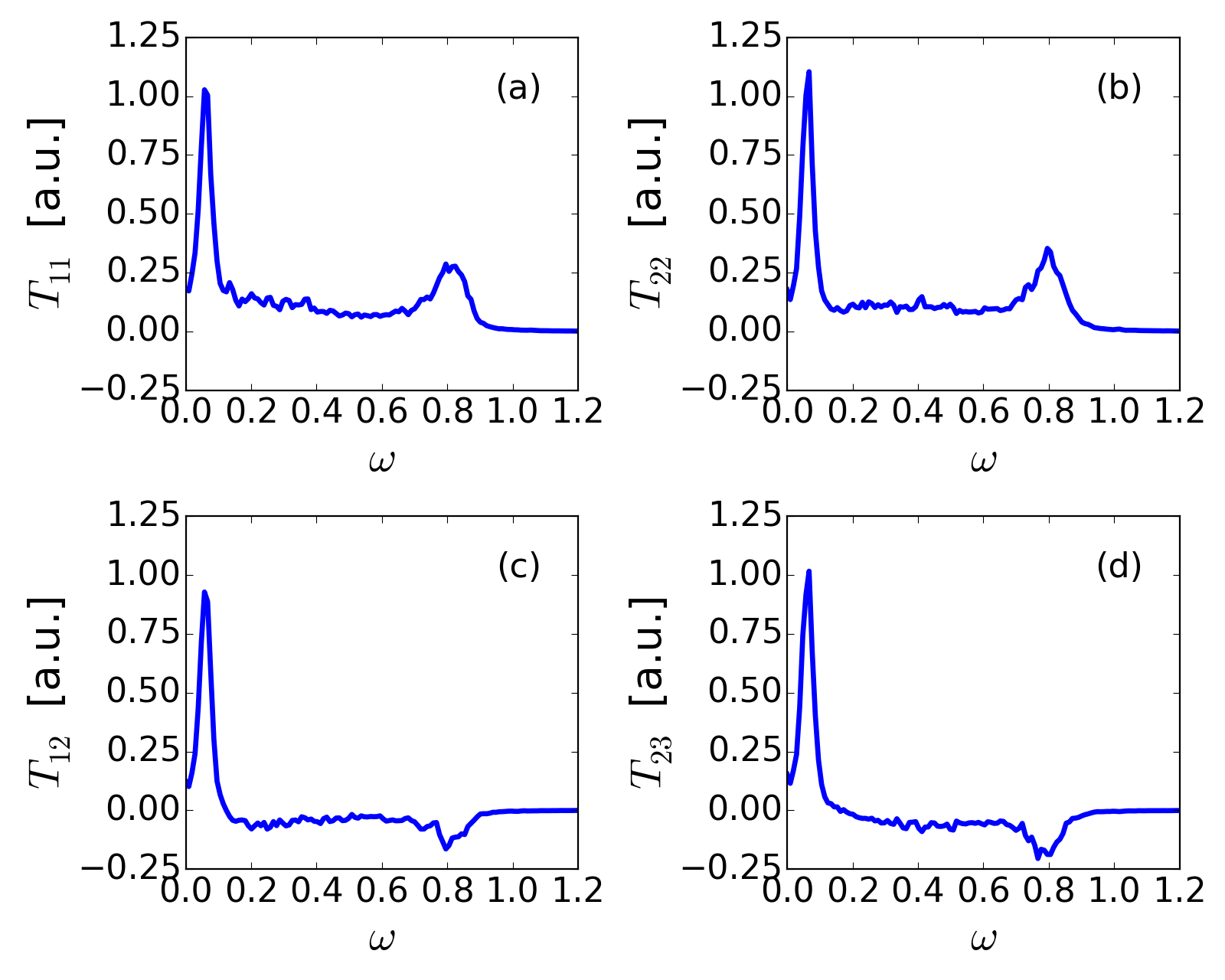}
\caption{The same as Fig.~\ref{fig:Lsd02} for the transverse current fluctuation spectra $T_{11}$, $T_{22}$, $T_{12}$, $T_{23}$. }
\label{fig:Tsd02}
\end{center}
\end{figure}

In this Section, we describe and interpret the formation of the envelope, a novel structure that emerges at low layer separations in the current fluctuation spectra and replaces the peaks which are associated with the gapped modes. 

\begin{figure}[htbp]
\begin{center}
\includegraphics[width=\columnwidth]{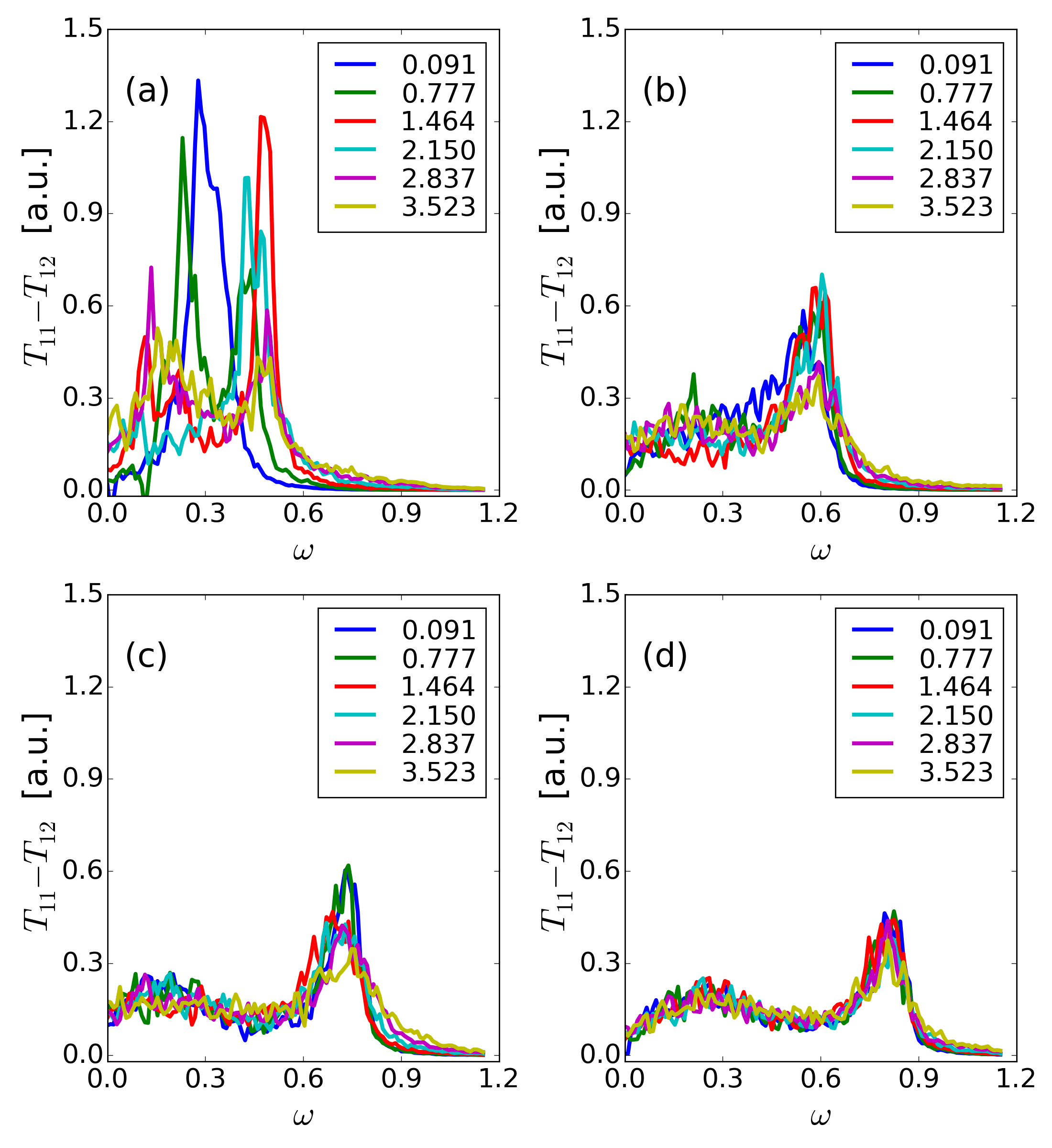}
\caption{The evolution of $T_{11}-T_{12}$ for $\Gamma$=160 with decreasing $d$ values. (a) $d=1.5$, (b) $d=1.0$, (c) $d=0.5$, (d) $d=0.2$. Note the transmutation of the $\mathcal{M}$ mode peak into the envelope.}
\label{fig:traT}
\end{center}
\end{figure}

\begin{figure}[htbp]
\begin{center}
\includegraphics[width=\columnwidth]{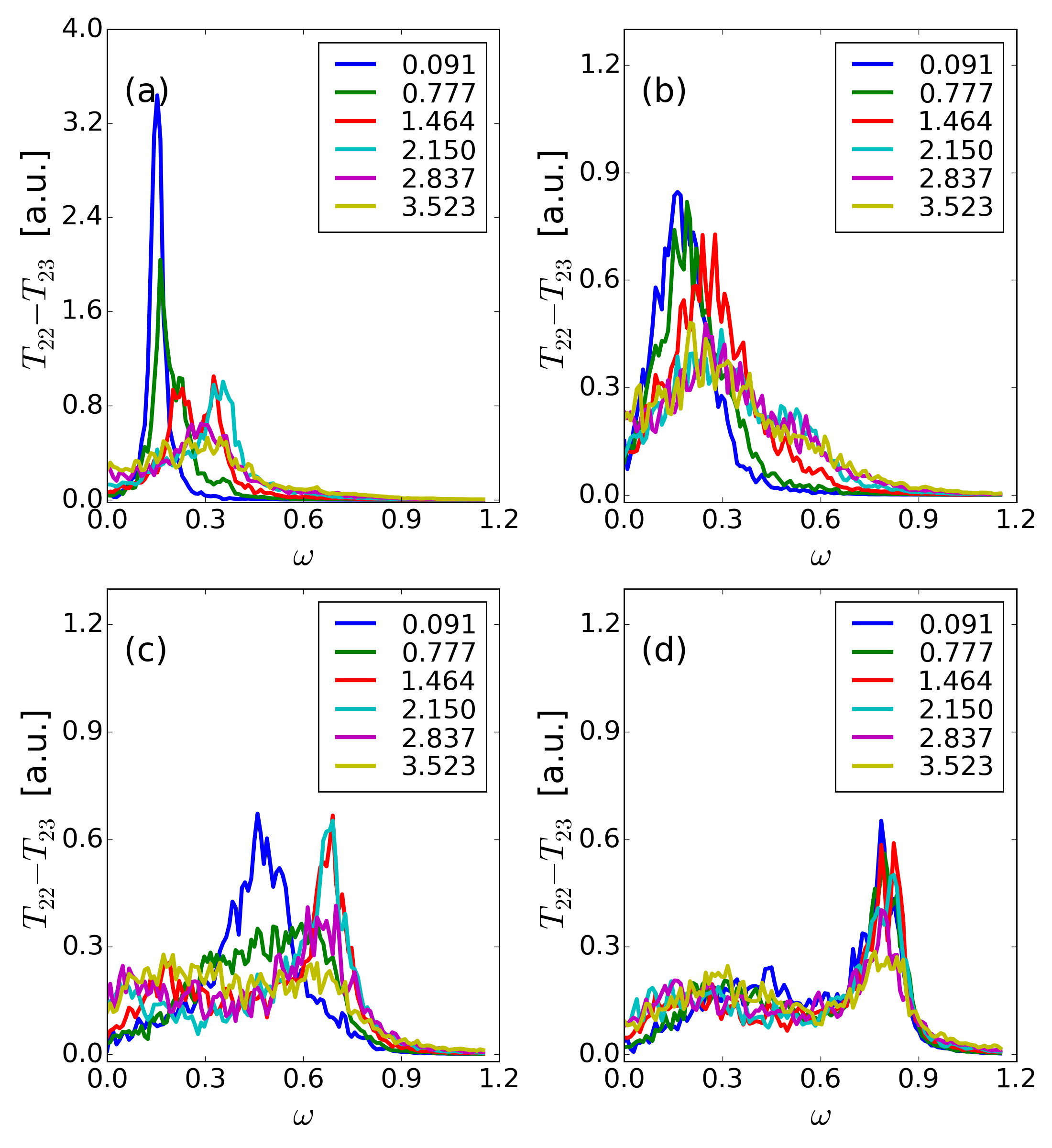}
\caption{ The same as Fig.~\ref{fig:traT} for $T_{22}-T_{23}$  Note the transmutation of the $\mathcal{S}$ mode peak into the envelope.}
\label{fig:traT23}
\end{center}
\end{figure}

A typical set of the current fluctuation spectra for the low $d=0.2$ value is displayed in Figs.~\ref{fig:Lsd02} and \ref{fig:Tsd02}. The observed spectral pattern is significantly different from the one seen in Fig.~\ref{fig:Lalls} for large inter-layer separations. The spectra now consist of two major domains. First, there are the high amplitude peaks representing the $\mathcal{AL}$ and $\mathcal{AT}$ acoustic modes where all the layers oscillate in-phase. As discussed in Section~\ref{sec:QLCA}, as $d \rightarrow 0$ these modes morph into the respective acoustic excitations of a 2D layer and are of no interest for the purpose of this Section. Second, we observe that the  peaks that would correspond to the QLCA predicted gapped modes are absent. Instead, the new feature to be noted is the appearance of the smaller positive (for the intra-layer correlations) or negative (for the inter-layer correlations) bumps, emerging beyond the domain of the QLCA gap excitations ($\omega \approx 0.6$). These features constitute the main body of the structure referred to as the ``envelope''. 
In order to focus now on the process of the disappearance of the gapped modes we will study the previously introduced differential spectra ($T_{11}-T_{12}$, etc.), where the acoustic excitations are absent. The evolution of the fluctuation spectra as the inter-layer distance is reduced can now be traced through the sequence of MD generated graphs for the range $d=1.5$ through $d=0.2$, presented in Figs.~\ref{fig:traT} and \ref{fig:traT23}.

From these and the previous set of data the main features of the envelope can be summarized as follows:
\begin{enumerate}
  \item The envelope has similar intra-layer structures in all the  $L_{AA}/T_{AA}$ and similar inter-layer structures in all the $L_{AB}/T_{AB}$ correlation functions. The respective structures become identical in the $d\to 0$ limit. This is expected, because the envelope dominated domain is governed by the substitutional disorder.
  \item In the intra-layer $L_{AA}/T_{AA}$-s the envelope necessarily has a positive value, but in the inter-layer $L_{AB}/T_{AB}$-s it always assumes a negative value. The intra-layer amplitudes are nearly twice (exactly twice in the $d\to 0$ limit) of the inter-layer amplitudes. These features ensure that the envelope does not show up in the total 2D $L$ and $T$ current fluctuation spectra, since $L_{\rm total} = \frac{1}{3} \left[L_{11}+L_{22}+L_{33}+2Re(L_{12}+L_{13}+L_{23})\right]$, which has the  well-known $\mathcal{AL}$ peak only.
  \item The amplitude of the envelope is much smaller than that of the $\mathcal{A}$ mode.
  \item The shape has a characteristic double-peaked feature. For high $\Gamma$, it exhibits a low amplitude peak at a low frequency $\omega\approx 0.3$ and a high amplitude peak at a high frequency at $\omega \approx 0.8$. For low $\Gamma$ it shows a central $\omega=0$ peak and a high frequency $\omega \approx 0.8$ peak.
  \item At the high $\Gamma$ value, the range, the amplitude and the shape of the envelope are all grossly insensitive to $k$ over a broad range of $k$ values.
\end{enumerate}

All these features are consistent with the observation discussed in Section~\ref{sec:structure}, noting that as $d \to 0$ substitutional disorder renders the layers indistinguishable. Remarkably, they also point at a possible relationship between the envelope and the velocity auto-correlation function (VAF) $Z(\omega)$~\cite{Schmidt1997}, where $Z(\omega)$ is the Fourier transform of $Z(t)$
\begin{equation} \label{eq:vaf}
  Z(t)=\langle\vec{v}_i(t) \vec{v}_i(0)\rangle.
\end{equation}

\begin{figure}[htbp]
\begin{center}
  \includegraphics[width=\columnwidth]{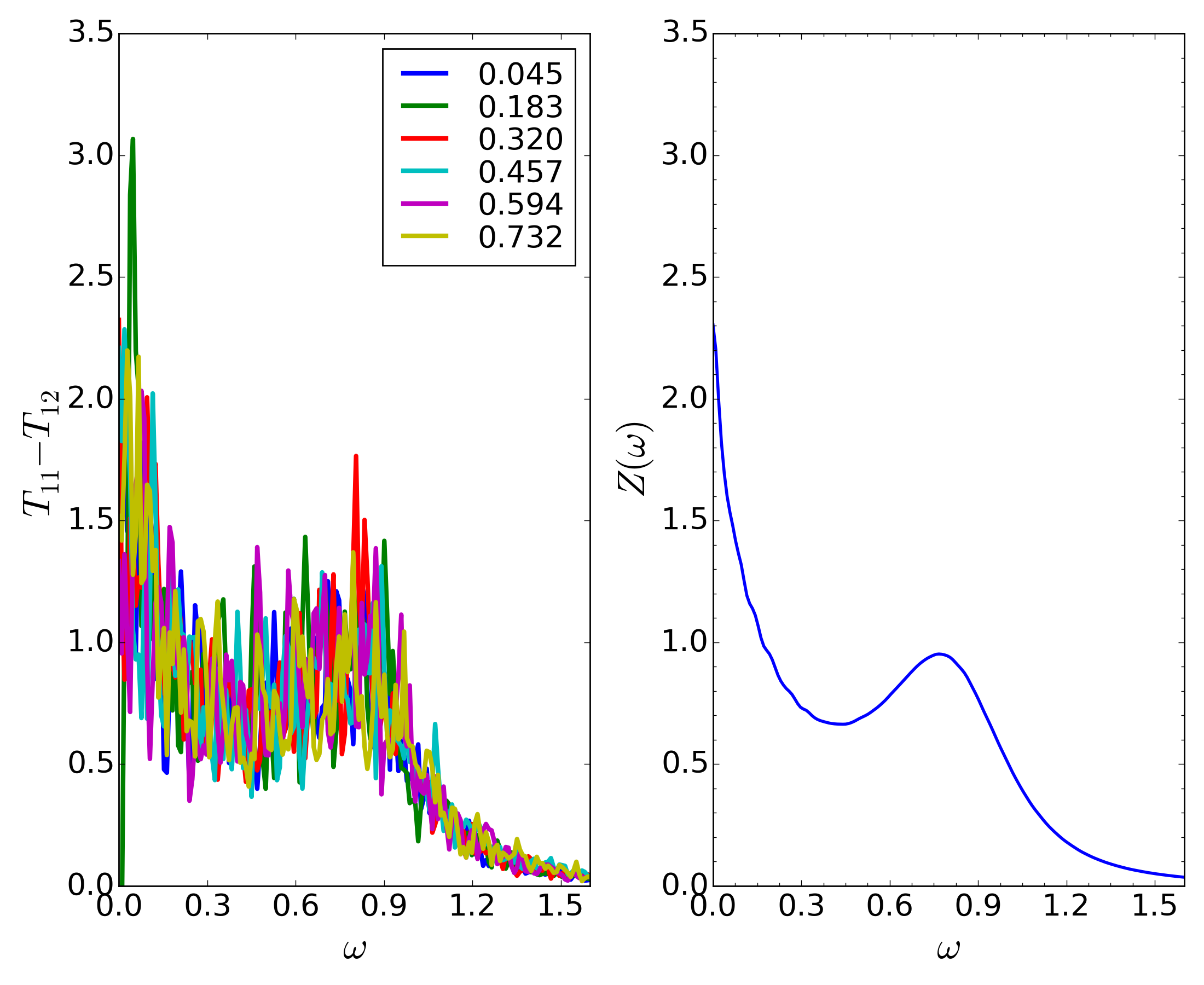}
  \caption{Comparison between $T_{11}-T_{12}$ and $Z(\omega)$ for $\Gamma$=10, at $d=0$.}
\label{fig:vafg10}
\end{center}
\end{figure}

\begin{figure}[htbp]
\begin{center}
  \includegraphics[width=\columnwidth]{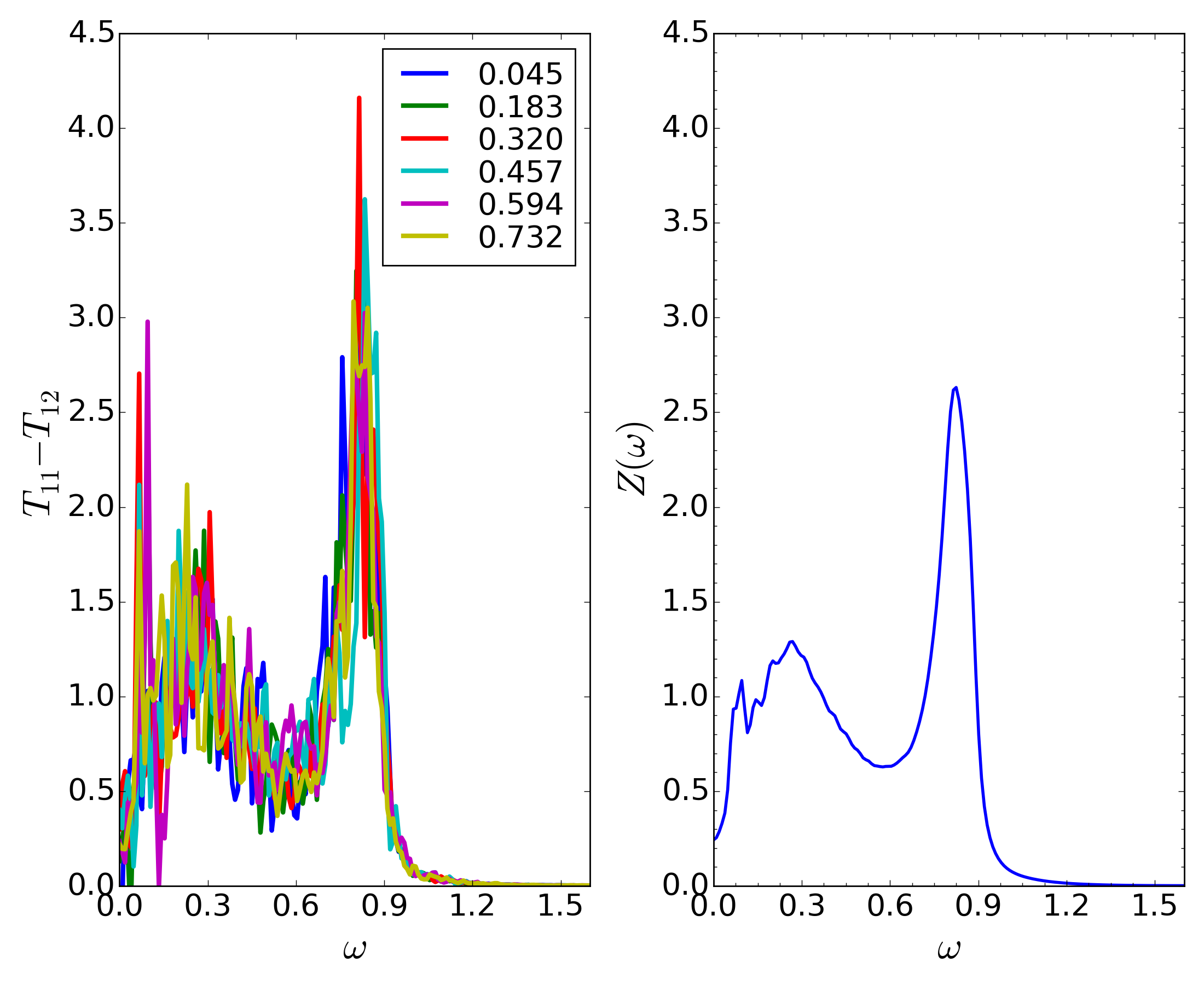}
  \caption{The same as Fig.~\ref{fig:vafg10} for $\Gamma$=160.}
\label{fig:vafg160}
\end{center}
\end{figure}

\begin{figure}[htbp]
\begin{center}
  \includegraphics[width=0.9\columnwidth]{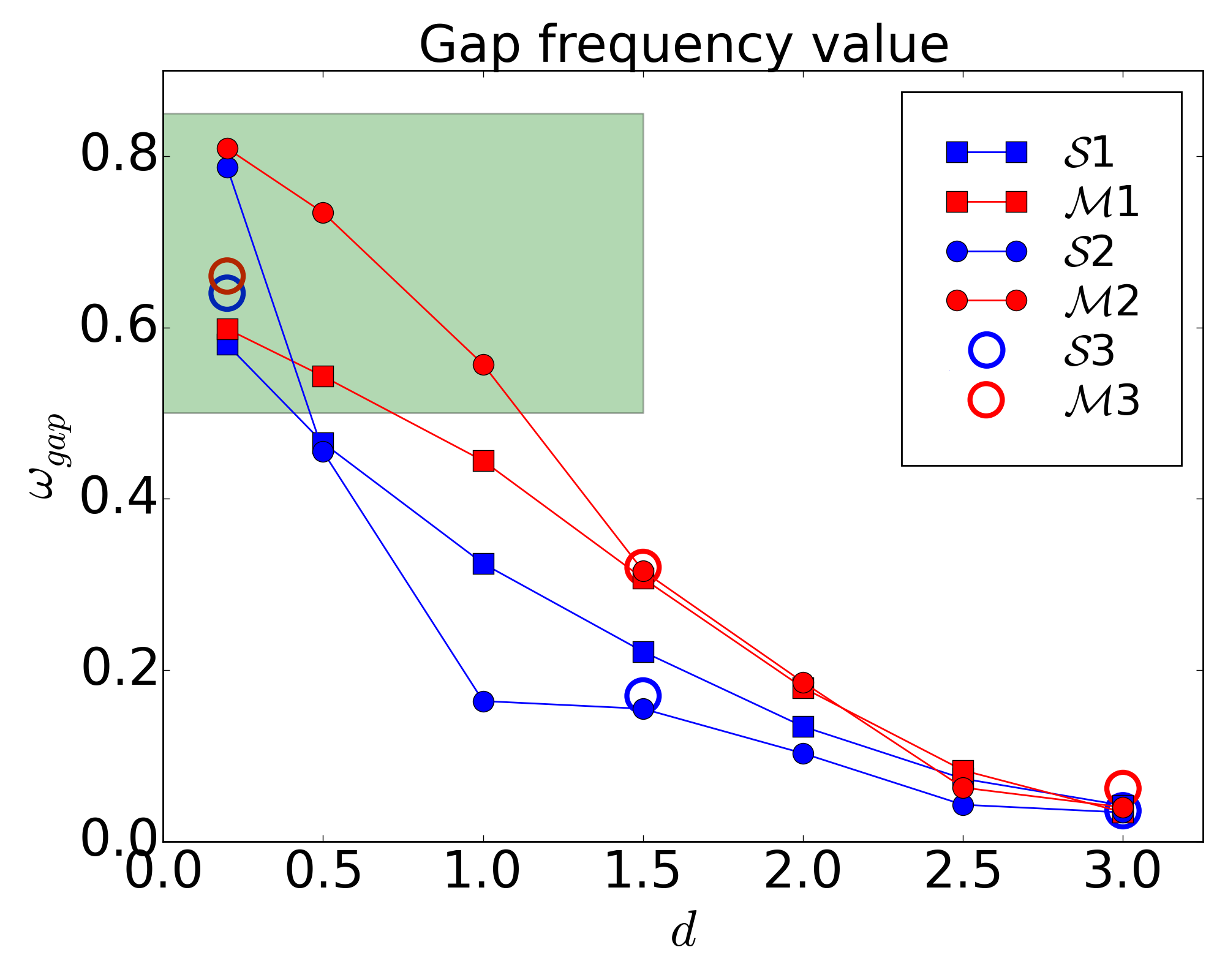}
  \caption{Gap frequency values for the $\mathcal{S}$ and $\mathcal{M}$ modes  at different $d$ values, for $\Gamma=160$. The S1, M1 points  are obtained from the QLCA theory (with the implicit assumption that the gapped modes are  excited for all $d$ values.); the S2, M2 points are are the results of the MD simulations; the S3, M3 points  are obtained from lattice dispersion calculations; at $d=3.0$ and at $d=0.2$ a SH, at $d=1.5$ an OS crystal structure is assumed (cf. Section III). The $\mathcal{S}2$ and $\mathcal{M}2$ points in the shaded area  originate from  peaks in the current fluctuation spectra generated by the  envelope, rather than by a collective mode.}
\label{fig:gapv}
\end{center}
\end{figure}

To see this more clearly, we compare the MD generated VAF-s for the trilayer system with those of the envelope. The same features can be found both in the longitudinal and in the  transverse spectra, here we use the transverse fluctuation spectrum.

For the sake of simplicity, the comparison is done in the $d=0$ limit. At this point some digression on this limiting process may be useful. As $d\to 0$, i.e. as the system becomes a 2D layer, all the features of the spectra extant at $d \ll 1$ survive. In the absence of substitutional disorder that would mean the survival, in addition to the acoustic modes, of the out-of-phase modes as well. With substitutional disorder, it would mean the continued presence of the envelope. On the other hand, neither of these features are observed in the spectrum of a 2D layer. This apparent paradox led reference~\cite{Ortner1999} to the erroneous conclusion that no gapped excitations can exist in a layered system in the liquid state. In fact, the resolution of the problem lies in the realization that the unexpected structures show up only in the partial (not total -- see earlier comment) fluctuation spectra. As long as $d$ is finite, groups of particles in different layers are distinguishable and therefore these structures represent measurable quantities. At $d=0$, when the groups are not distinguishable, they do not; they are useful mathematical constructs, nevertheless.

Two sets of spectra are displayed in Figs.~\ref{fig:vafg10} and \ref{fig:vafg160}, both for low and high $\Gamma$ values. The apparent agreement between the envelope and the VAF functions suggests a strong connection between them. Even the diffusive peak at $\omega \to 0$ that appears in the VAF in the low $\Gamma$ case is present in the envelope. Even though one may suspect that the connection with the VAF (a one-body entity) suggests an $N$-dependent single particle rather than a collective behavior, we have verified in separate computations that the envelope profile is independent of the particle number used in the MD simulation.
 
While a cogent theoretical explanation for the link between the envelope and the VAF is lacking at this time, the conclusion for the existence of such a relationship, presumably via the onset of substitutional disorder at low $d$ values, seems to be well-founded. Moreover, preliminary computational tests on a perfect ordered lattice have shown that an envelope-like structure forms, when an artificially induced substitutional disorder is generated. Also, further investigations of a bilayer system at small $d$ values have confirmed a behavior very similar to the one found here, with the appearance of the same kind of an envelope. All this requires further investigation and clarification and will be discussed in a separate publication.

Finally, the evolution of the gap frequency as a function of the layer separation is portrayed in Fig.~\ref{fig:gapv}. The QLCA curve in the diagram is based on Eq.~(\ref{eq:gapvs}). In addition to the MD simulation data, a few frequency values obtained from calculations through the harmonic phonon approximation for the respective SH and OS ideal lattice structures, appropriate for the layer distances indicated, are also given. The results of the lattice calculations and those of the QLCA are quite close to each other in most situations, except at very low layer separations where the QLCA values are substantially lower. The reason for the discrepancy should be sought in the fact that the QLCA calculation based on the actual PDF reflects the effect of the substitutional disorder, while the lattice dispersion, based on the idealized crystal lattice model, does not. The MD data are in good agreement with the calculated values for higher $d$-values, but not in  the shaded small $d$-domain,  for reasons discussed in this Section. As $d\to 0$ the MD points for both modes converge to $\omega=0.8$, the characteristic peak value of the envelope, while the QLCA gap frequency equals the Einstein frequency of a projected 2D layer at $\omega=0.6$.

In previous papers on strongly coupled Coulomb bilayer liquids~\cite{Kalman1999,Donko2003a,Donko2003}  the high frequency peak of the envelope was erroneously identified to be the $d \rightarrow 0$ continuation of the peak associated with the gapped collective mode of the bilayer (cf. Fig.~\ref{fig:gapv}). This led to the incorrect conclusion that the gapped mode survives all the way to $d=0$, but with a sizable discrepancy between the MD results and the QLCA predictions, resulting in a gap frequency at about 1.4 times higher than the predicted value. In fact, as can be gleaned from Fig.~\ref{fig:gapv}, a similar gap discrepancy could be arrived at for the trilayer as well through an incorrect inference from the data points. The correct interpretation of the data has to be sought along the line presented in the foregoing discussions.


\section{\label{sec:conclusion} Conclusion}

In this paper we have studied the equilibrium structure and the dynamics of the collective excitations of a Yukawa trilayer in the strongly coupled liquid state. The study has been done for $\kappa=0.4$ screening parameter value. Both analytic and MD simulation techniques have been used, the former based on the QLCA method. The equilibrium structure has been found to reveal a variety of liquid phases, depending on the distance between the layers. These phases emulate the different underlying ground state structures that would form in a trilayer in the solid state at the same layer separation. In this respect, the trilayer is similar to the bilayer, but for an additional degree of freedom that allows for two possible relative configurations of the top and bottom layers: they can be either shifted with respect to each other (ABC stacking) or be in an overlapping configuration (ABA stacking). In the latter case an effective attraction, mediated by the middle layer,  between these two layers is observed. At smaller ($d<1$) inter-layer distances the system develops a substitutionally disordered state, in which particles in a given layer of the underlying lattice occupy the ``wrong'' lattice position belonging to a neighboring layer. 

In the collective mode spectrum we have identified six modes distinguished by their Cartesian polarizations (longitudinal or transverse) and by the relative movements of the layers, which can be either in-phase with an acoustic dispersion, or out-of-phase with an optic (gapped) dispersion. There are two gapped modes, associated with two gap frequencies, corresponding to two different relative movements of the layers. In the dispersion of these modes we find a notable manifestation of the ``avoided crossing'' phenomenon, known mostly in other physical contexts.
At the relatively low, down to $\Gamma=50$ value there is a rather surprising presence of solid microcrystals, witnessed by the appearance of the imprint of anisotropic lattice modes in the collective  mode spectrum. At larger inter-layer distances, the analytic QLCA predictions match the MD results  well. However, when the inter-layer distance drops below $d\approx 1$, the nascent substitutional disorder then quenches the gapped modes, which vanish from the excitation spectrum. Instead, a novel structure emerges in all the current fluctuation spectra, covering a broad frequency range that extends beyond the typical frequencies of the gapped modes. We have referred to this new feature as the ``envelope''. While its origin is not well understood at this time, it seems to be related to the developing substitutional disorder. The morphology of the envelope strongly suggests that it derives from the velocity auto-correlation function of the system, whose spectral form it closely mirrors.

In this study we have identified a few  issues, whose impact on the behavior of layered systems  in general may be of relevance, meriting further future investigations: 

\begin{itemize}
    \item the formation of a striped equilibrium phase,
	\item the development of substitutional disorder,
	\item the condition for the appearance of the avoided crossing points in the mode dispersions,
	\item the formation of microcrystals; their effect on the mode spectrum in the liquid phase,
    \item the ``envelope'' phenomenon; the role of the Velocity Autocorrelation Function in the fluctuation spectra.
\end{itemize}

\begin{acknowledgments}

The authors thank Luciano Silvestri for discussions and comments. The work has been partially supported by by the National Science Foundation under Grant PHY-1613102 and PHY-1740203. Z. D. and P. H. gratefully acknowledge financial support from the Hungarian Office for Research, Development and Innovation under NKFIH grants K-119357 and K-132158.

\end{acknowledgments}

\onecolumngrid
 
\appendix

\section{C-matrix elements}

In the following formulae $C_{\mu\nu}^{AB}$ is given in units of $\omega_{\rm p}^2$.  $\rho$ is the projected 2D distance, $r\equiv r_{AB}$ is the full 3D distance; they are connected by $r^2 = \rho^2+s_{AB}^2$, where $s_{AB}$ is the distance between layers $A$ and $B$. The $AB$ subscripts match the indices of the $g_{AB}$ PDF-s they are linked to.
Note that $P$ and $Q$ stand for $P\equiv P(\kappa r)$, $Q\equiv Q(\kappa r)$. 

\begin{align}
    \begin{split}
    C^{11}_{xx}&=\int_0^\infty\frac{{\rm e}^{-\kappa r}}{3r^5}\left(\frac{1}{2}P\rho^2-Qr^2\right)g_{11}(\rho)\rho {\rm d} \rho  \\
    &+\int_0^\infty\frac{{\rm e}^{-\kappa r}}{3r^5}\left(\frac{1}{2}P\rho^2-Qr^2\right)g_{12}(\rho)\rho {\rm d} \rho  \\
    &+\int_0^\infty\frac{{\rm e}^{-\kappa r}}{3r^5}\left(\frac{1}{2}P\rho^2-Qr^2\right)g_{13}(\rho)\rho {\rm d} \rho  \\
    &-\int_0^\infty\frac{{\rm e}^{-\kappa r}}{3r^5}\left[\left(\frac{J_1(k\rho)}{k\rho}-J_2(k\rho)\right)P\rho^2-Qr^2J_0(k\rho)\right]g_{11}(\rho)\rho {\rm d} \rho 
    \end{split}
\end{align}

\begin{align}
    \begin{split}
    C^{22}_{xx}&=\int_0^\infty\frac{{\rm e}^{-\kappa r}}{3r^5}\left(\frac{1}{2}P\rho^2-Qr^2\right)g_{21}(\rho)\rho {\rm d} \rho  \\
    &+\int_0^\infty\frac{{\rm e}^{-\kappa r}}{3r^5}\left(\frac{1}{2}P\rho^2-Qr^2\right)g_{22}(\rho)\rho {\rm d} \rho  \\
    &+\int_0^\infty\frac{{\rm e}^{-\kappa r}}{3r^5}\left(\frac{1}{2}P\rho^2-Qr^2\right)g_{23}(\rho)\rho {\rm d} \rho  \\
    &-\int_0^\infty\frac{{\rm e}^{-\kappa r}}{3r^5}\left[\left(\frac{J_1(k\rho)}{k\rho}-J_2(k\rho)\right)P\rho^2-Qr^2J_0(k\rho)\right]g_{22}(\rho)\rho {\rm d} \rho 
    \end{split}
\end{align}

\begin{align}
    \begin{split}
    C^{12}_{xx}=
    &-\int_0^\infty\frac{{\rm e}^{-\kappa r}}{3r^5}\left[\left(\frac{J_1(k\rho)}{k\rho}-J_2(k\rho)\right)P\rho^2-Qr^2J_0(k\rho)\right]g_{12}(\rho)\rho {\rm d} \rho 
    \end{split}
\end{align}
  
\begin{align}
    \begin{split}
    C^{23}_{xx}=
    &-\int_0^\infty\frac{{\rm e}^{-\kappa r}}{3r^5}\left[\left(\frac{J_1(k\rho)}{k\rho}-J_2(k\rho)\right)P\rho^2-Qr^2J_0(k\rho)\right]g_{23}(\rho)\rho {\rm d} \rho 
    \end{split}
\end{align}

\begin{align}
    \begin{split}
    C^{11}_{yy}&=\int_0^\infty\frac{{\rm e}^{-\kappa r}}{3r^5}\left(\frac{1}{2}P\rho^2-Qr^2\right)g_{11}(\rho)\rho {\rm d} \rho \\
    &+\int_0^\infty\frac{{\rm e}^{-\kappa r}}{3r^5}\left(\frac{1}{2}P\rho^2-Qr^2\right)g_{12}(\rho)\rho {\rm d} \rho \\
    &+\int_0^\infty\frac{{\rm e}^{-\kappa r}}{3r^5}\left(\frac{1}{2}P\rho^2-Qr^2\right)g_{13}(\rho)\rho {\rm d} \rho \\
    &-\int_0^\infty\frac{{\rm e}^{-\kappa r}}{3r^5}\left[\frac{J_1(k\rho)}{k\rho}P\rho^2-Qr^2J_0(k\rho)\right]g_{11}(\rho)\rho {\rm d} \rho 
    \end{split}
\end{align}

\begin{align}
    \begin{split}
    C^{22}_{yy}&=\int_0^\infty\frac{{\rm e}^{-\kappa r}}{3r^5}\left(\frac{1}{2}P\rho^2-Qr^2\right)g_{21}(\rho)\rho {\rm d} \rho \\
    &+\int_0^\infty\frac{{\rm e}^{-\kappa r}}{3r^5}\left(\frac{1}{2}P\rho^2-Qr^2\right)g_{22}(\rho)\rho {\rm d} \rho \\
    &+\int_0^\infty\frac{{\rm e}^{-\kappa r}}{3r^5}\left(\frac{1}{2}P\rho^2-Qr^2\right)g_{23}(\rho)\rho {\rm d} \rho \\
    &-\int_0^\infty\frac{{\rm e}^{-\kappa r}}{3r^5}\left[\frac{J_1(k\rho)}{k\rho}P\rho^2-Qr^2J_0(k\rho)\right]g_{22}(\rho)\rho {\rm d} \rho 
    \end{split}
\end{align}

\begin{align}
    \begin{split}
    C^{12}_{yy}&=-\int_0^\infty\frac{{\rm e}^{-\kappa r}}{3r^5}\left[\frac{J_1(k\rho)}{k\rho}P\rho^2-Qr^2J_0(k\rho)\right]g_{12}(\rho)\rho {\rm d} \rho 
    \end{split}
\end{align}

\begin{align}
    \begin{split}
   C^{23}_{yy}&=-\int_0^\infty\frac{{\rm e}^{-\kappa r}}{3r^5}\left[\frac{J_1(k\rho)}{k\rho}P\rho^2-Qr^2J_0(k\rho)\right]g_{23}(\rho)\rho {\rm d} \rho 
    \end{split} 
\end{align}

\twocolumngrid
\bibliography{references}

\end{document}